\documentclass[pra,showpacs,showkeys,superscriptaddress,amssymb,amsmath]{revtex4}
\usepackage{graphicx}
\usepackage{hyperref}
\usepackage{placeins}
\usepackage{xcolor}
\usepackage{caption}
\usepackage{listings}
\usepackage[normalem]{ulem}
\usepackage{color}
\usepackage{float}
\usepackage{simplewick}

\begin{document}

	\title{
		A recursive enumeration of connected Feynman diagrams with an arbitrary number of external legs in the fermionic non-relativistic interacting gas
	}
	
	\author{E. R. Castro}
	\email[]{erickc@cbpf.br}
	\affiliation{Centro Brasileiro de Pesquisas F\'{\i}sicas/MCTI,
		22290-180, Rio de Janeiro, RJ, Brazil}	
	
	\author{I. Roditi}
	\email[]{roditi@cbpf.br}
	\affiliation{Centro Brasileiro de Pesquisas F\'{\i}sicas/MCTI,
		22290-180, Rio de Janeiro, RJ, Brazil}

	\begin{abstract}
		In this work, we generalize a recursive enumerative formula for connected Feynman diagrams with two external legs. The Feynman diagrams are defined from a fermionic gas with a two-body interaction. The generalized recurrence is valid for connected Feynman diagrams with an arbitrary number of external legs and an arbitrary order. The recurrence formula terms are expressed in function of weak compositions of non-negative integers and partitions of positive integers in such a way that to each term of the recurrence correspond a partition and a weak composition. The foundation of this enumeration is the Wick theorem, permitting an easy generalization to any quantum field theory. The iterative enumeration is constructive and enables a fast computation of the number of connected Feynman diagrams for a large amount of cases.  In particular, the recurrence is solved exactly for two and four external legs, leading to the asymptotic expansion of the number of different connected Feynman diagrams.    
	\end{abstract}	
	\keywords{Connected Feynman diagrams, Counting Feynman diagrams, Non-relativistic interaction gas, Asymptotics methods, Enumerative combinatorics, Zero-dimensional field theory, Wick theorem.}
	\pacs{02.10.Ox; 02.30.Mv; 31.15.xp}
	\maketitle

	\section{Introduction} \label{Int}
	
	Enumeration of Feynman diagrams is currently an active research subject in quantum field \cite{Cvitanovic}\cite{Kleinert}\cite{Brezin} and many-body theoretical research \cite{Molinari}\cite{Molinari2}\cite{Pavlyukh}. The formal perturbative machinery flows into well-defined operations that unambiguously define the Feynman diagrams. The combinatorial character of this generative process is contained in two equivalent formalisms: the functional and the field operator approaches. The diagrams represent processes expressed commonly in terms of divergent integrals whose contribution is obtained afterwards by renormalization. Although the enumeration of the Feynman diagrams is independent of the integrals that represent the physical processes, when we take the total contribution of certain classes of diagrams, the global structure of the generative combinatorics is relevant. (This can be seen, for instance, in recent results\cite{Cavalcanti}, where the symmetry factor -or multiplicity- of the related Feynman diagrams appear explicitly in the integrals.)
	
	The standard way to count Feynman diagrams is to define the theory in zero dimension within the QFT functional approach\cite{Cvitanovic}\cite{Molinari}\cite{Argyres}. (Here, the functional integral is transformed into a conventional integral.) The zero-dimensional theory can be understood as a toy model for the study of the formal mathematical machinery used in non-zero dimension quantum field theory\cite{Rivasseau}\cite{Borinsky2018Graphs}. In particular, as a simplified model, exact results are possible and it could be extendable to non-zero-dimensional field theory.
	
	Recently ref.\cite{Castro1} used a different principle for counting Feynman diagrams: in well-defined algebraic (multiplicative) relations between objects expressed as sums of Feynman diagrams associated with each object, the replacement of the sums by the explicit number of total contractions that generates the specific diagrams in each order leads to recursive relations between the number of total contractions associated with each object for each order of perturbation. In the generating function terminology, this principle has straightforward interpretation. For example, consider the following ordinary multiplicative relation between the arbitrary generating functions $\mathcal{G}_{A}(g)$, $\mathcal{G}_{B}(g)$, $\mathcal{G}_{C}(g)$ and $\mathcal{G}_{D}(g)$, 

\begin{equation}\label{G1}\mathcal{G}_{A}(g)=\mathcal{G}_{B}(g)\times\mathcal{G}_{C}(g)\times\mathcal{G}_{D}(g),\end{equation} where $g$ is the arbitrary parameter in which the generating function would be expanded in formal series. The function $\mathcal{G}_{X}(g)$ can be a correlation function, an $n$-point function etc, each one expressed as a sum of diagrams for all the perturbation orders (in the many body case, for example, eq. \ref{G1} can be induced from the Dyson equations, the indices $A,B,\cdots,X$ are used only to distinguish between this generating functions). $g$ can be the coupling constant of the theory. In zero dimension, these functions lose their explicit dependence on the space-time coordinates and take the form of the following formal series in $g$:

\begin{equation}\label{G}\mathcal{G}_{X}(g)=\sum_{m=0}^{\infty}\mathfrak{N}_{m}^{(X)}g^{m},
\end{equation}
where $\mathfrak{N}_{m}^{(X)}$ is the number of $m$-order Wick total contractions present in $\mathcal{G}_{X}(g)$. Expression (\ref{G}) is the generating function of the number of $m$-order total contractions and it induces in (\ref{G1}) the following sum, which associates the respective number of contractions for each order

\begin{equation}\mathfrak{N}_{m}^{(A)}=\sum_{m_{1}=0}^{m}\sum_{m_{2}=0}^{m}\sum_{m_{3}=0}^{m}\delta_{m_{1}+m_{2}+m_{3},m}\mathfrak{N}_{m_{1}}^{(B)}\mathfrak{N}_{m_{2}}^{(C)}\mathfrak{N}_{m_{3}}^{(D)}.
\end{equation}
	
 When the associated $m$-order Feynman diagrams have the same multiplicity, these relations determine the number of different Feynman diagrams.
 
 \begin{figure}[H]
	\centering
		\includegraphics[width=0.13\textwidth , angle=270]{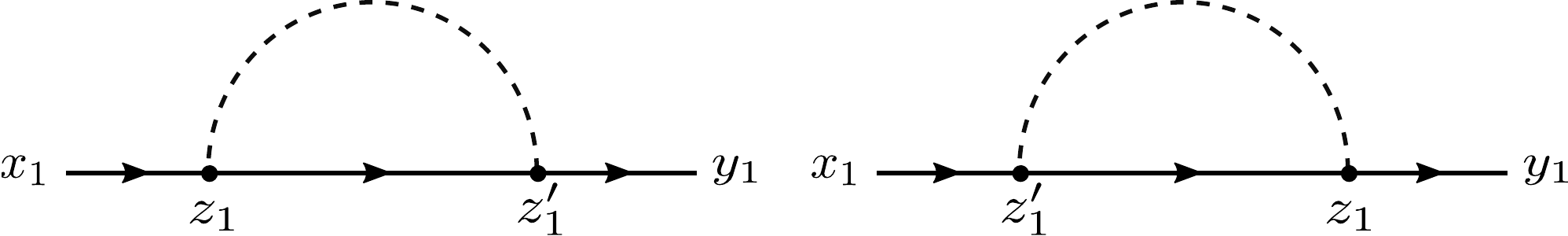}
	\caption{A total contraction corresponds with a precise rule to draw a Feynman diagram, The left diagram is generated by the rule $(x_{1}\to z_{1})(z_{1}\to z_{1}^{\prime})(z_{1}^{\prime}\to y_{1})$. The right diagram is generated by the rule $(x_{1}\to z_{1}^{\prime})(z_{1}^{\prime}\to z_{1})(z_{1}\to y_{1})$. The two rules are different, however they generate the same drawning (Feynman diagram). The multiciplity of the corresponding Feynman diagram is two.}
	\label{Multi}
\end{figure}
 
If all the $m$-order Feynman diagrams have the same multiplicity $M$,  then the number of different diagrams is simply $\mathfrak{N}_{m}^{(A)}/M$, where $\mathfrak{N}_{m}$ is the number of $m$-order total contractions. Particularly, this is the case in QED and in many-body theory for connected Feynman diagrams with external legs. (Vacuum Feynman diagrams, and disconnected Feynman diagrams with vacuum components do not satisfy this rule.)This simple counting principle was also used by Ref.\cite{Kugler} in a generalized way to determine the number of different types of Feynman diagrams from certain many-body relations, leading to an efficient counting of a great variety of Feynman diagrams (Hugenholtz diagrams, bare Feynman diagrams, skeletons Feynman diagrams, etc.) There is vast literature dedicated to the counting of Feynman diagrams. See Ref.\cite{Castro2} for a brief introduction and, for an exhaustive study, see Refs\cite{Borinsky2018Graphs},\cite{Borinsky} and \cite{Argyres}.  
	
	In this work, we generalize the previous recursive enumerative formula for connected Feynman diagrams with two external legs for a fermionic interacting gas\cite{Castro1} to the case of connected Feynman diagrams with an arbitrary number of external legs. The recursive enumerative formula was used in Ref.\cite{Castro1} to get an exact formula and find an equivalence with the Arqu\'es-B\'eraud formula for one-rooted maps (i.e., objects in algebraic topology)\cite{Prunotto}. Particularly, equivalences between the counting of $N$-rooted maps and connected Feynman diagrams with $2N$ external legs have been established by means of a directed bijection between these two types of object\cite{Gopala2}\cite{Gopala}. Exact formulas related to this algebraic curve topological theory have also been obtained, which, can also be used to count Feynman diagrams. Other connections between Feynman diagrams and rooted maps can be seen in Ref.\cite{Yeats}. A Rooted map is a graph that is embedded in a unique topological surface (sphere or $n$-holed tori) with a directed edge.
	
	\begin{figure}[H]
	\centering
		\includegraphics[width=0.18\textwidth , angle=270]{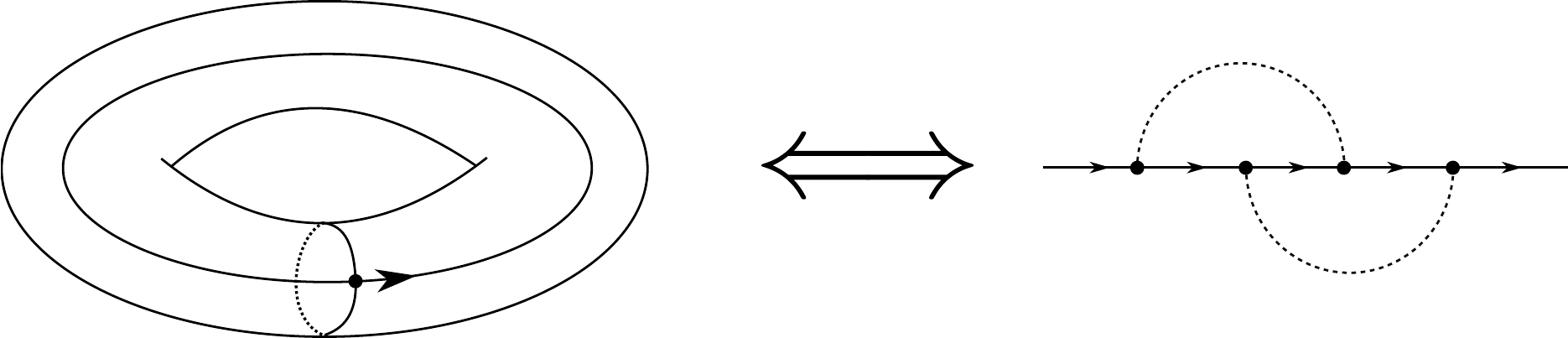}
	\caption{A rooted map embedded in the tori and the corresponding Feynman diagram. See all the correspondences for order $m=1,2,3$ in ref. \cite{Prunotto}}
	\label{Rmf}
\end{figure}
	
	The derivation of our recurrence formula start with the Wick Theorem and has a possible interpretation in terms of elementary combinatorial theory. Based on the bijection found in Ref.\cite{Gopala2}, our counting also applies to the $N$-rooted map case and can be considered a different enumerative process.
	
	This paper is organized as follows. In section \ref{W}, from the set of possible Wick contractions, we establish the possible ways to construct an arbitrary disconnected Feynman diagram. By summing over all the possibilities, we obtain a set of recurrences, which relate the number of connected Feynman diagrams for different orders and the number of external legs. In section \ref{R}, we enormously simplify the recurrence, reducing it to a form that allows an easy computation of the number of different connected Feynman diagrams. Section \ref{E} exactly solves the recurrences for $N=1$ and $N=2$ (two and four external legs, respectively) and from these exact values we find many terms of the respective asymptotic expansions. We also we determine the existence of a new type of asymptotic contribution (negligible in respect to the main contribution) which is not present in the conventional literature. Section \ref{D} contains the final considerations of this work.  
	
\section{Wick Theorem and factorization in terms of connected Feynman diagrams} \label{W} 
	  
	Let us consider the Hamiltonian $\hat{H}=\hat{H}_{0}+\hat{H}_{I}$, where $\hat{H}_{0}$ is the free Hamiltonian containing the free kinetics terms and $\hat{H}_{I}$ is the interaction Hamiltonian containing all the two-body interactions. The object that generates the $m$-order Feynman diagrams with two external legs is the following expectation value in the free ground state $|\phi_{0}\rangle$
	 
	 \begin{equation}\label{Interac}
	 \langle\phi_{0}|T[\hat{H}_{I}(t_{1})\cdots\hat{H}_{I}(t_{m})\hat{\psi}_{\alpha}(x)\hat{\psi}_{\beta}^{\dagger}(y)]|\phi_{0}\rangle,
	 \end{equation}
     where $x$ and $y$ are the external and fixed space-time variables, $T[\cdots]$ is the time ordering product of $[\cdots]$, $\alpha$ and $\beta$ are the spinor indices of the respective field operators, and $\hat{H}_{I}(t_{i})$ is the interaction Hamiltonian in the interaction picture scheme. In particular $\hat{H}_{I}(t_{i})$ in the second-quantization form is

\begin{equation}\label{HI}
\hat{H}_{I}(t_{i})=\frac{1}{2}\sum_{\lambda_{i}\lambda_{i}^{\prime} \mu_{i}\mu_{i}^{\prime}}\int d^{3}\vec{z_{i}}d^{3}\vec{z_{i}^{\prime}}\hat{\psi}_{\lambda_{i}^{\prime}}^{\dagger}(z_{i})\hat{\psi}_{\mu_{i}^{\prime}}^{\dagger}(z_{i}^{\prime})U(z_{i},z_{i}^{\prime})_{\lambda_{i}\lambda_{i}^{\prime} \mu_{i}\mu_{i}^{\prime}}\hat{\psi}_{\lambda_{i}}(z_{i})\hat{\psi}_{\mu_{i}}(z_{i}^{\prime}),
\end{equation}
	where $z_{i}=(\vec{z_{i}},t_{i})$ and $z_{i}^{\prime}=(\vec{z_{i}^{\prime}},t_{i})$ are the internal space-time variables, $i \in \{1,2,\cdots,m\}$ is the index associated to the interaction $U(z_{i},z_{i}^{\prime})$. Considering this interaction as of the Coulomb type, the associated system would be a non-relativistic interacting gas of identical particles. The indices $\lambda_{i},\lambda_{i}^{\prime},\mu_{i}$ and $\mu_{i}^{\prime}$ are spinorial indices. In $U(z_{i},z_{i}^{\prime})$, the indices express the possibility of spin interaction between the particles. Note that all the variables (spinor indices and space time variables) related to the internal vertices are added or integrated, meaning that (\ref{Interac}) are the coefficients of a $2\times 2$ matrix indexed by $\alpha$ and $\beta$. In the fermionic case, the precise rules for the construction of the Feynman diagram are given, for example, in chapter 3, section 9 of Ref.\cite{Fetter}. We will only consider the Feynman diagrams in the fermionic case. The bosonic case in the many body context is different in zero temperature. However, for finite temperature our approach is valid in the fermion and bosonic case, see chapter 7 of ref.\cite{Fetter}. The $2m$ variables $\{z_{i},z_{i}^{\prime}\}$ correspond to the $2m$ internal vertices. The interaction $U(z_{i},z_{i}^{\prime})$ corresponds to a dashed line edge joining the vertices $z_{i}$ and $z_{i}^{\prime}$, The diagram order is given respect to the number of this interaction-dashed lines. The fermionic directed leg starting at $z_{a}$ and ending in $z_{b}$ corresponds to the Wick contraction association 
	
	\begin{equation}
\contraction{\hat{\psi}^{\dagger}(z_{1})\cdots}{\hat{\psi}}{{}^{\dagger}(z_{b})\cdots}{\hat{\psi}}	
	\hat{\psi}^{\dagger}(z_{1})\cdots\hat{\psi}^{\dagger}(z_{b})\cdots\hat{\psi}(z_{a})\cdots\hat{\psi}(z_{m}^{\prime})\hat{\psi}(x)\hat{\psi}^{\dagger}(y)=\overbrace{\hat{\psi}(z_{a})\hat{\psi}^{\dagger}(z_{b})}\hat{\psi}^{\dagger}(z_{1})\cdots\hat{\psi}(z_{m}^{\prime})\hat{\psi}(x)\hat{\psi}^{\dagger}(y).
	\end{equation} 
	
	The incoming (outcoming) external legs are obtained (respectively) by the substitution $z_{a} \to x$ ($z_{b} \to y$) in the previous expression. The respective Feynman diagrams are obtained by contracting all the field operators in all the possible ways (total contractions). The factor $$\overbrace{\hat{\psi}(z_{a})\hat{\psi}^{\dagger}(z_{b})}=\overbrace{\hat{\psi}^{\dagger}(z_{b})\hat{\psi}(z_{a})}$$ is a $c$-number (the free propagator) depending on $z_{a}$ and $z_{b}$ (and also on the respective spinor indices). After contracting, we can treat it as a simple number.
	
\subsection{Wick theorem for Feynman diagrams with arbitrary number of external legs}
	
	 These rules are easily generalized to the case of $2N$ external legs. In this case, the Feynman generator expectation value is 
	
	\begin{equation}\label{Contract}
	\langle\phi_{0}|T[\hat{H}_{I}(t_{1})\cdots\hat{H}_{I}(t_{m})\hat{\psi}_{\alpha_{1}}(x_{1})\hat{\psi}_{\alpha_{2}}(x_{2})\cdots\hat{\psi}_{\alpha_{N}}(x_{N})\hat{\psi}_{\beta_{1}}^{\dagger}(y_{1})\hat{\psi}_{\beta_{2}}^{\dagger}(y_{2})\cdots\hat{\psi}_{\beta_{N}}^{\dagger}(y_{N})]|\phi_{0}\rangle.
	\end{equation}
	
	From now on, for simplicity, we omit the spinor indices, which will be considered in each respective internal variable $z_{i}$ or in the external variables $x_{j}$ and $y_{k}$. Every total contraction (and the respective Feynman diagram) is actually a tensorial contribution to the $2N$ spinorial components of the external variables. The $2m$ internal spins components are summed. This can be observed in expression (\ref{HI}).
	
	For example, the diagram with four external legs
	$$\includegraphics[scale=.5]{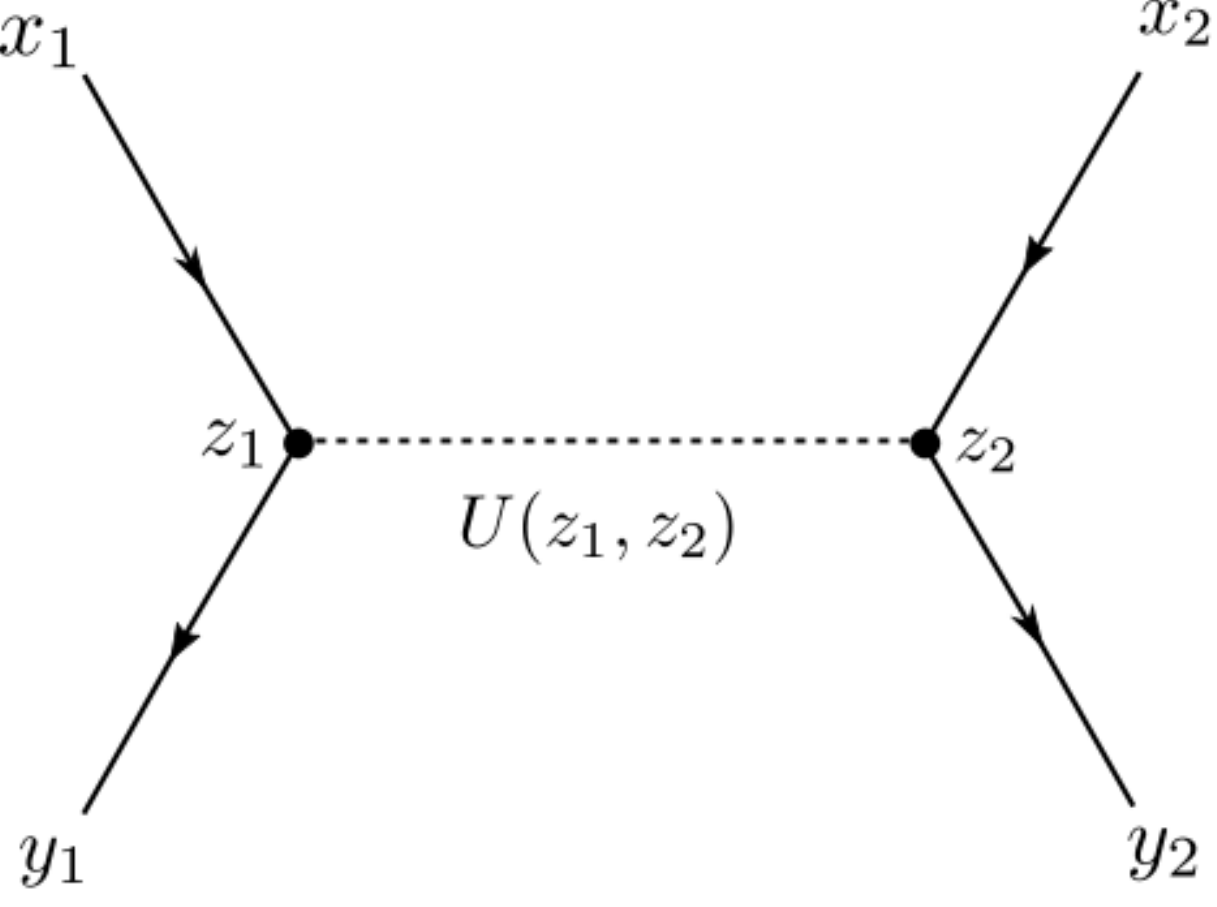}$$ corresponds with the total contraction
	
	$$\contraction{}{\hat{\psi}}{{}^{\dagger}(z_{1})\hat{\psi}^{\dagger}(z_{2})\hat{\psi}(z_{1})\hat{\psi}(z_{2})}{\hat{\psi}}
	\contraction[2ex]{\hat{\psi}^{\dagger}(z_{1})}{\hat{\psi}}{{}^{\dagger}(z_{2})\hat{\psi}(z_{1})\hat{\psi}(z_{2})\hat{\psi}(x_{1})}{\hat{\psi}}
	\bcontraction[3ex]{\hat{\psi}^{\dagger}(z_{1})\hat{\psi}^{\dagger}(z_{2})}{\hat{\psi}}{{}(z_{1})\hat{\psi}(z_{2})\hat{\psi}(x_{1})\hat{\psi}(x_{2})}{\hat{\psi}}
	\bcontraction[4ex]{\hat{\psi}^{\dagger}(z_{1})\hat{\psi}^{\dagger}(z_{2})\hat{\psi}(z_{1})}{\hat{\psi}}{{}(z_{2})\hat{\psi}(x_{1})\hat{\psi}(x_{2})\hat{\psi}^{\dagger}(y_{1})}{\hat{\psi}}
	\hat{\psi}^{\dagger}(z_{1})\hat{\psi}^{\dagger}(z_{2})\hat{\psi}(z_{1})\hat{\psi}(z_{2})\hat{\psi}(x_{1})\hat{\psi}(x_{2})\hat{\psi}^{\dagger}(y_{1})\hat{\psi}^{\dagger}(y_{2})U(z_{1},z_{2})$$
	
	or
 $$\overbrace{\hat{\psi}(x_{1})\hat{\psi}^{\dagger}(z_{1})}\times\overbrace{\hat{\psi}(z_{1})\hat{\psi}^{\dagger}(y_{1})}U(z_{1},z_{2})\overbrace{\hat{\psi}(z_{2})\hat{\psi}^{\dagger}(y_{2})}\times\overbrace{\hat{\psi}(x_{2})\hat{\psi}^{\dagger}(z_{2})}.$$	
	The number of possible total contractions (i.e., possible association in pairs) $\mathfrak{N}_{m}^{(N)}$ in (\ref{Contract}) is
	
	\begin{equation} \label{TotalCont}
    \mathfrak{N}_{m}^{(N)}=(2m+N)!.
	\end{equation}
	This is easy to see: In expression (\ref{Contract}), there are $2m+N$ annihilation field operators ($N$ different $\hat{\psi}(x_{i})$, $m$ different $\hat{\psi}(z_{j})$ and  $m$ different $\hat{\psi}(z_{k}^{\prime})$) and $2m+N$ creation field operators ($N$ different $\hat{\psi}^{\dagger}(y_{i})$, $m$ different $\hat{\psi}^{\dagger}(z_{j})$ and  $m$ different $\hat{\psi}^{\dagger}(z_{k}^{\prime})$.) Non-vanishing contractions only happen between creation and annihilation operators. Therefore, the total number of possible contractions is $(2m+N)!$. (All the field operators must be contracted.)
	
  Also, it is easy to see that the total number of external legs is always even. For each internal vertex $z_{b}$, there are only two associate field operators, $\hat{\psi}(z_{b})$ and $\hat{\psi}^{\dagger}(z_{b})$, which are contracted between them,
  
   \begin{equation}
  \includegraphics[scale=.6]{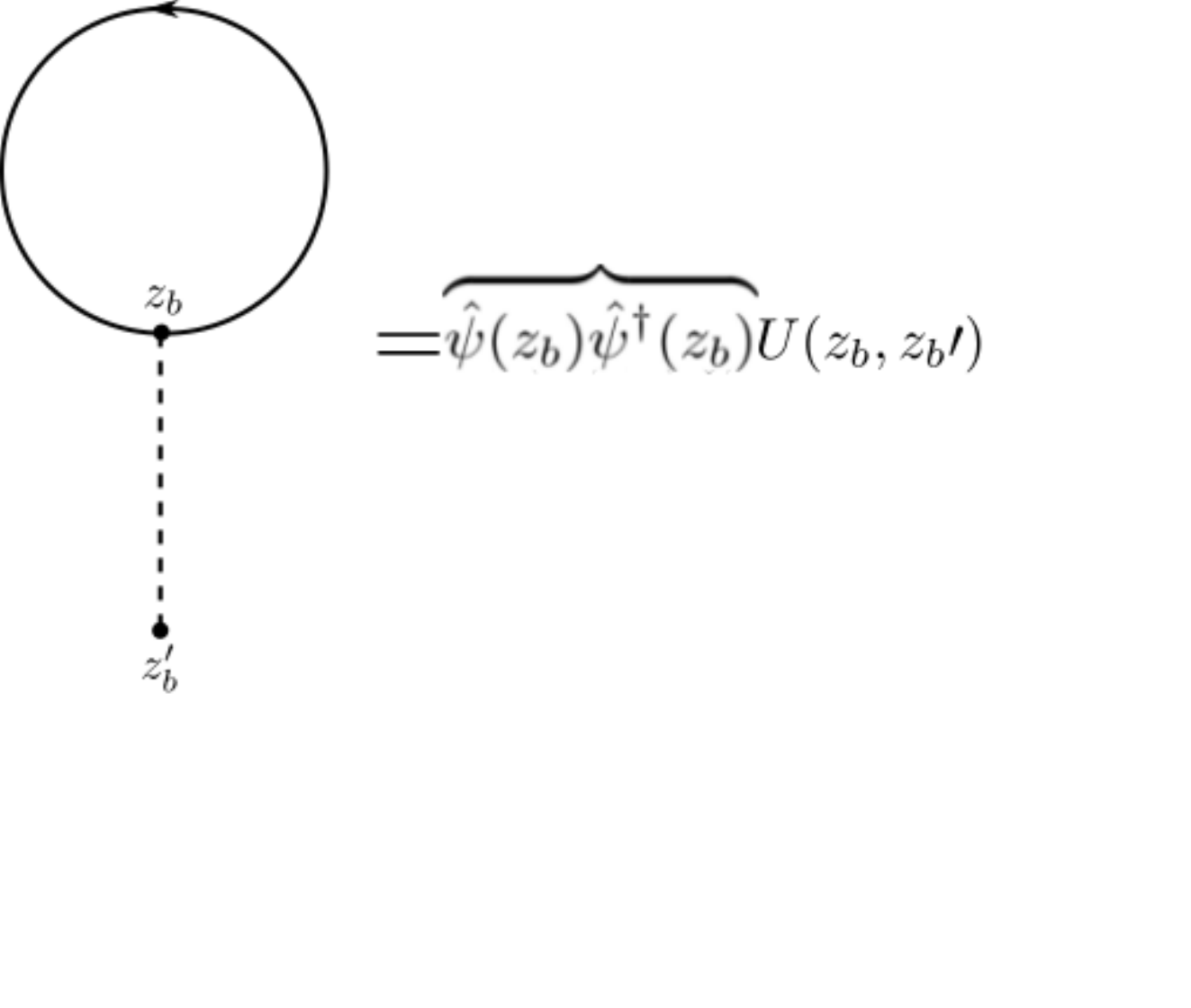}
  \end{equation} 
  or with $\hat{\psi}^{\dagger}(z_{c})$ and $\hat{\psi}(z_{a})$, respectively, 
  
\begin{equation}
  \includegraphics[scale=.7]{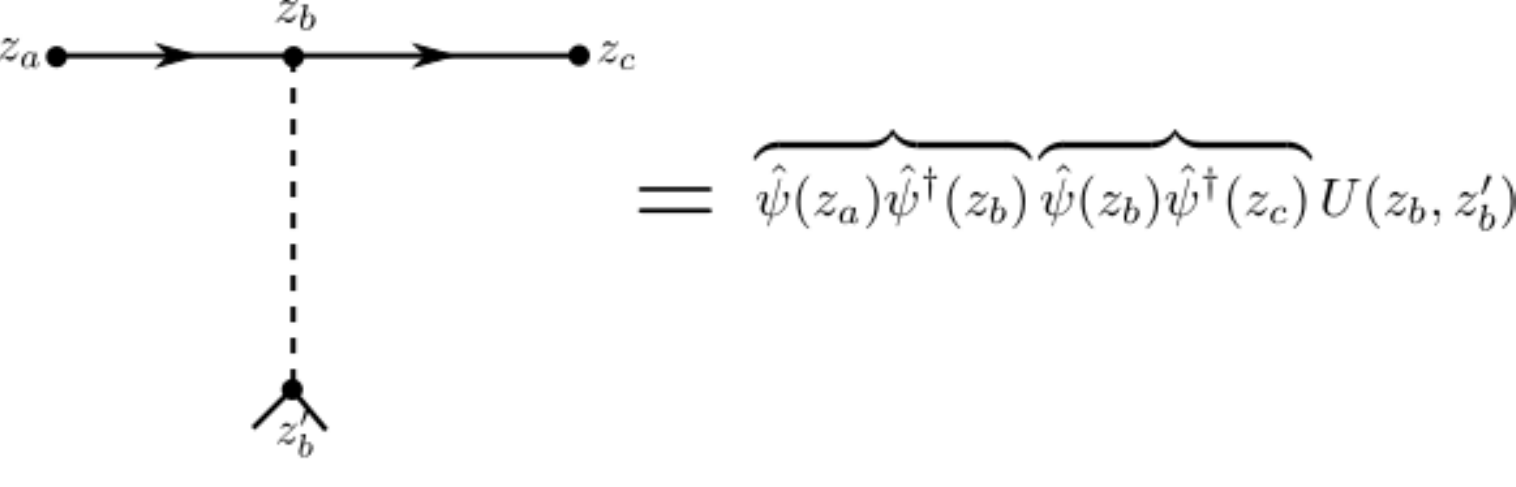}.
  \end{equation}
  Therefore, each vertex $z_{i}$ belongs to a unique trail of fermion lines that is a closed cycle (with one or more internal vertices) or is a trail which begins in a unique $x_{a}$ and ends in a unique $y_{b}$. (The only field operators associated with $x_{a}$ and $y_{b}$ are $\hat{\psi}(x_{a})$ and $\hat{\psi}^{\dagger}(y_{b})$, respectively). An odd number of external legs implies the existence of at least one vertex with more than two associated field operators, which is a contradiction. 
  
  The dashed line $U(z_{j},z_{j}^{\prime})$ connects the vertices $z_{j}$ and $z_{j}^{\prime}$, which may belong to the same fermionic trail or to different trails. For a given contraction, we have a set of trails and, by inserting the fixed interactions $U(z_{j},z_{j}^{\prime})$, we obtain the corresponding Feynman diagram. This diagram can be connected or disconnected.
  
\subsection{General decomposition of a $\mathbf{m}$-order disconnected Feynman diagram and factorization property of the diagrams with the same decomposition}  
  
   In order to find a formula for the total number of possible contractions, we consider an arbitrary contraction of order $m$ with $2N$ external legs, and then we add all the possibilities. Suppose an arbitrary $m$-order disconnected Feynman diagram $\mathcal{F}=\mathcal{F}_{1}\times\mathcal{F}_{2}\times\cdots\times\mathcal{F}_{l+1} $     
  
	\begin{equation}
	\includegraphics[scale=.7, angle=270]{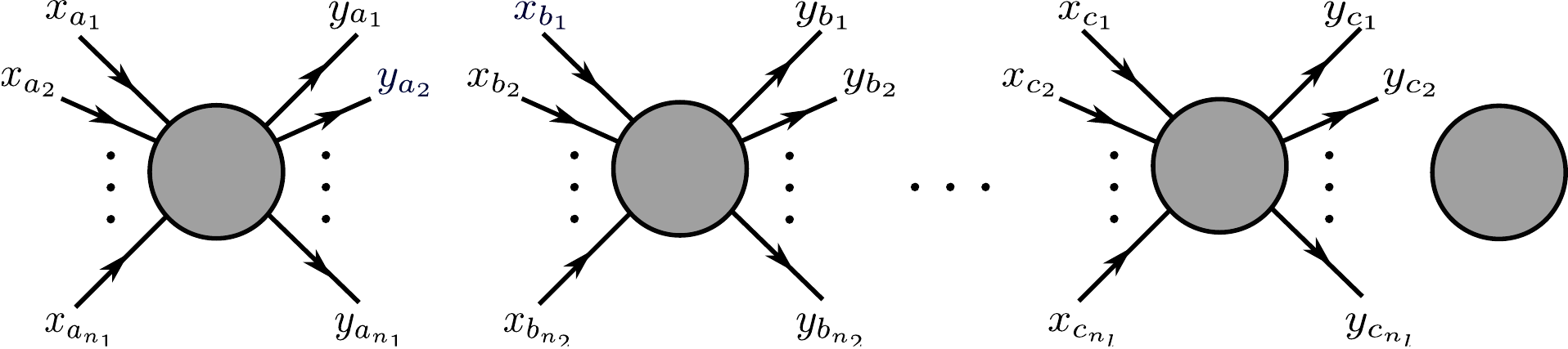} \label{graph}
	\end{equation}
	which has $l$ connected components, each of them with order $m_{1},m_{2},\cdots m_{l}$, respectively. (The $m_{l}$ interacting dashed lines are inside the gray disc of the $l$ connected component.) The vertical sequence of dots expresses that for every component we have a sequence of incoming (outgoing) external legs (the first component have $2n_{1}$ external legs, the second have $2n_{2}$ and so on). The horizontal sequence of dots expresses the sequence of connected components of an arbitrary disconnected Feynman diagram, which have respective orders $m_{1},m_{2},\cdots,m_{l}$. The last component is not necessarily connected, its order is $m_{l+1}$, and it only contains vacuum-bubble diagrams. 
	
	We have, then \begin{equation}m=m_{1}+m_{2}+\cdots+m_{l+1}\label{order}\end{equation} and \begin{equation}N=n_{1}+n_{2}+\cdots+n_{l}.\label{legs}\end{equation} In how many ways can we choose the internal vertices of each component? The order of the first component is $m_{1}$, so there are $\binom{m}{m_{1}}$ ways to choose the pairs $(z_{i},z_{i}^{\prime})$. The order of the second component is $m_{2}$, and we then have $\binom{m-m_{1}}{m_{2}}$ ways to choose the pairs, and so on. Thereby, the number of possible choices is
	
	\begin{equation}
	\binom{m}{m_{1}}\binom{m-m_{1}}{m_{2}}\cdots\binom{m-m_{1}-\cdots-m_{l}}{m_{l+1}}= \frac{m!}{m_{1}!m_{2}!\cdots m_{l+1}!}.
	\end{equation}
	
	The external legs can also be chosen in different ways. There are $N!$ ways to choose the incoming $N$ legs. As we are not yet investigating the internal structure of each component, for now, it only matters to know the different forms to associate the outgoing legs with each component. The first component has $n_{1}$ outgoing legs. Therefore, the number of possibilities in choosing the outgoing legs in the first component, once the incoming legs are fixed, is $\binom{N}{n_{1}}$. For the second component, there exist $\binom{N-n_{1}}{n_{2}}$ possibilities, and so on. Once the incoming lines are fixed, the total number of possibilities is
	
	\begin{equation}
	\binom{N}{n_{1}}\binom{N-n_{1}}{n_{2}}\cdots\binom{N-n_{1}-\cdots-n_{l-1}}{n_{l}}= \frac{N!}{n_{1}!n_{2}!\cdots n_{l}!}.\label{Multinom}
	\end{equation}
	
\begin{figure}
	\centering
		\includegraphics[width=0.7\textwidth]{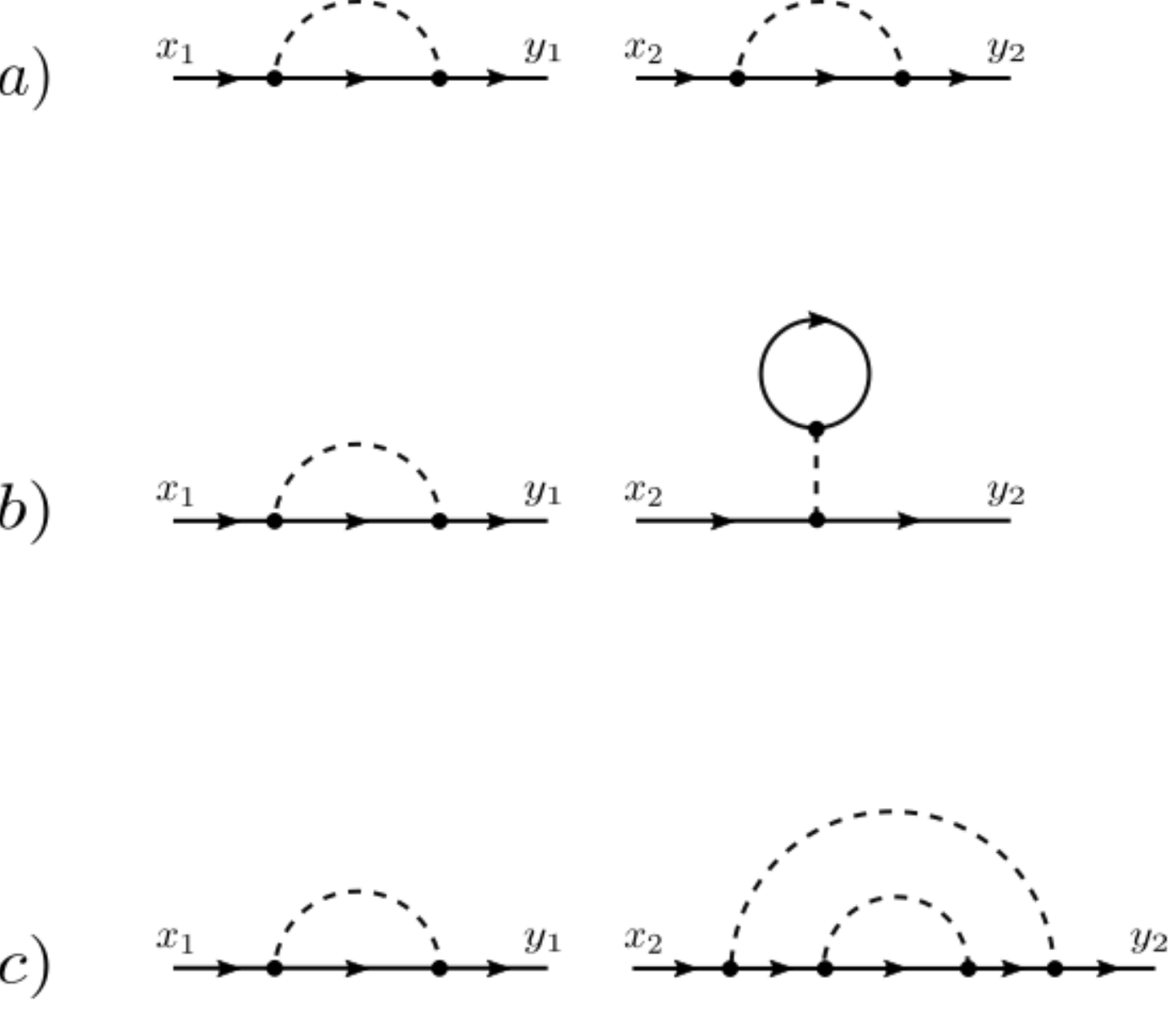}
	\caption{(a) Disconnected diagram with two connected components and with $m=2$ and $N=2$. The four possible enumerations of the external legs are $\{(x_{1},y_{1})(x_{2},y_{2}),(x_{1},y_{2})(x_{2},y_{1}),(x_{2},y_{1})(x_{1},y_{2}),(x_{2},y_{2})(x_{1},y_{1})\}$. Notice that there are only two different enumerations, given the identical components. The counting is $(N!)^{2}/(n_{1}!n_{2}!d_{1}!)=2$ with $N=2, n_{1}=n_{2}=1$ and $d_{1}=2$. (b) In the second case, the two components are different, and we have four different possible enumerations of the external legs: the first two are given by the factor $(N!)^{2}/(n_{1}!n_{2}!d_{1}!)=2$. When we consider all the contractions $\mathcal{N}_{c\,m_{1}}$ and $\mathcal{N}_{c\,m_{2}}$ in the product $\mathcal{N}_{c\,m_{1}}\mathcal{N}_{c\,m_{2}}$ with $m_{1}=m_{2}=1$, we have the case where the different components of (b) are exchanged, this gives the others two. (c) In the third case, the disconnected Feynman diagram has order $m=3$, with $m_{1}=1$ and $m_{2}=2$. The two components are different, and we have four possible enumerations of the external legs. The counting factor $(N!)^{2}/(n_{1}!n_{2}!d_{1}!)=2$ gives 2 belonging to $\mathcal{N}_{c\,1}\mathcal{N}_{c\,2}$. The other two contribution happens when we take $m_{1}=2$ and $m_{2}=1$ and the components are exchanged in $\mathcal{N}_{c\,2}\mathcal{N}_{c\,1}$. Without the factor $d_{i}$ We would have over-counting in all the cases.}
	\label{OverCounting}
\end{figure}

	If we have a different number of external legs for each component, the total number of possibilities is simply $N!$ multiplied by the multinomial coefficient expressed in (\ref{Multinom}). This is an over-counting if there are components with equal number of external legs, see fig \ref{OverCounting}. In particular, if we have only $r<l$ components with different number of external legs such that $N=d_{1}n_{1}+d_{2}n_{2}+\cdots+d_{r}n_{r}$, where $d_{i}$ is the number of components with the same number of external legs (and, evidently, $l=d_{1}+\cdots+d_{r}$), the correct counting, in this case, is given by
	
	\begin{equation}
	 \frac{1}{d_{1}!d_{2}!\cdots d_{r}!}\times\frac{(N!)^{2}}{n_{1}!n_{2}!\cdots n_{l}!}.\label{Count}
	\end{equation}
	
	Now, it is time to study the internal structure of each component. Note that all the Feynman diagrams that satisfy (\ref{order}) and (\ref{legs}), have the same external structure and, therefore, they carry the same counting as in (\ref{Count}). (I.e., the substitution of the component $h$ by another with the same order $m_{h}$ and the same number $2n_{h}$ of external legs leads to the same counting as expressed in (\ref{Count}).) Bearing that the diagram in (\ref{graph}) represents a product of $l+1$ integrals, it follows that the sum of all the different diagram contributions that satisfy (\ref{order}) and (\ref{legs}) is factored in a product whose $l+1$ elements are the sum of all the possible components. The internal structure is considered by taking, instead of all the possible different components, the different contractions that have the respective order and number of external legs in each component. So, the total number of possible contractions satisfying (\ref{order}) and (\ref{legs}) is
	
	\begin{equation}
\frac{1}{d_{1}!d_{2}!\cdots d_{r}!}\times\frac{(N!)^{2}}{n_{1}!n_{2}!\cdots n_{l}!}\times\frac{m!}{m_{1}!m_{2}!\cdots m_{l+1}!}\mathcal{T}[\mathcal{F}_{1}]_{n_{1}}^{m_{1}}\times\cdots\times\mathcal{T}[\mathcal{F}_{l}]_{n_{l}}^{m_{l}}\times\mathcal{T}[\mathcal{F}_{l+1}]_{0}^{m_{l+1}}
	\end{equation}
	with
	
	\begin{equation}
	\mathcal{T}[\mathcal{F}_{i}]_{n_{i}}^{m_{i}}=\langle\phi_{0}|T[\hat{H}_{I}(t_{1})\cdots\hat{H}_{I}(t_{m_{i}})\hat{\psi}(x_{e_{1}})\hat{\psi}(x_{e_{2}})\cdots\hat{\psi}(x_{e_{n_{i}}})\hat{\psi}^{\dagger}(y_{e_{1}})\hat{\psi}^{\dagger}(y_{e_{2}})\cdots\hat{\psi}^{\dagger}(y_{e_{n_{i}}})]|\phi_{0}\rangle_{connected},
	\end{equation}
	where $i \in \{1,2,\cdots,l\}$, and
	
	\begin{equation}
	\mathcal{T}[\mathcal{F}_{l+1}]_{0}^{m_{l+1}}=\langle\phi_{0}|T[\hat{H}_{I}(t_{1})\cdots\hat{H}_{I}(t_{m_{l+1}})|\phi_{0}\rangle.
	\end{equation}
	
	The index $connected$ implies that we are only considering contractions that generate connected diagrams. Suppose that there is a total of $\mathcal{N}_{c\,m_{i}}^{(n_{i})}$ of such contractions. Let us replace $\mathcal{T}[\mathcal{F}_{i}]_{n_{i}}^{m_{i}}$ by $\mathcal{N}_{c\,m_{i}}^{(n_{i})}$ and $\mathcal{T}[\mathcal{F}_{l+1}]_{0}^{m_{l+1}}$ by the number of all possible contractions $\mathfrak{D}_{m_{l+1}}$ that generate vacuum-bubble Feynman diagrams (connected and disconnected). This number is obtained making $N=0$ in (\ref{TotalCont}) 
	\begin{equation}\label{Vacuum}
	\mathfrak{D}_{m_{l+1}}=(2m_{l+1})!.
	\end{equation} 
	
	Therefore, the total number of contractions that generate Feynman diagrams satisfying (\ref{order}) and (\ref{legs}) is: 
	
\begin{equation}
\frac{1}{d_{1}!d_{2}!\cdots d_{r}!}\times\frac{(N!)^{2}}{n_{1}!n_{2}!\cdots n_{l}!}\times\frac{m!}{m_{1}!m_{2}!\cdots m_{l+1}!}\mathcal{N}_{c\,m_{1}}^{(n_{1})}\mathcal{N}_{c\, m_{2}}^{(n_{2})}\cdots \mathcal{N}_{c\,m_{l}}^{(n_{l})}\mathfrak{D}_{m_{l+1}}.\label{Particular}
\end{equation}

One important consideration: In formula (\ref{Particular}) we count, in the multiplicative factor, all the possible ways to choose the external legs. (Therefore, when we speak of the $\mathcal{N}_{c\,m}^{(N)}$ connected total contractions the choice of external legs has already been made in the corresponding diagrams.) From now on, when we speak of Feynman diagrams, the choice of external legs will be the following: the first fermion trail will begin in $x_{1}$ and  end in $y_{1}$,  the second fermion trail will begin in $x_{2}$ and  end in $y_{2}$, and so on.

\subsubsection{The least possible order of a connected diagram with $2N$ external legs}

Before continuing, let us verify what the minimal order possible of a connected diagram with $2N$ legs is $m=N-1$. For N=1, the diagram with minimal order possible is evidently the free propagator, with m=0. For $N=2$, the minimal order connected Feynman diagram is $m=1$:

\begin{equation}\label{Feyn}
\includegraphics[scale=.5]{4Legs.pdf}. 
\end{equation}

In diagram (\ref{Feyn}), we have two trails, each one with an internal vertex, and one dashed line connecting these internal vertices. For $N=3$, we use the previous case to build the minimal order connected diagram. We must add another trail with two external legs and one internal vertex. To connect this trail, we simply add one internal vertex to one of the above trails and connect it to one additional dashed line. This construction is the minimal possible. So, for $N=3$ we have $m=2$. This construction can be generalized for all the other cases, obtaining $m=N-1$ as the mimimal possible order.

\begin{figure}[H]
	\centering
		\includegraphics[width=0.7\textwidth]{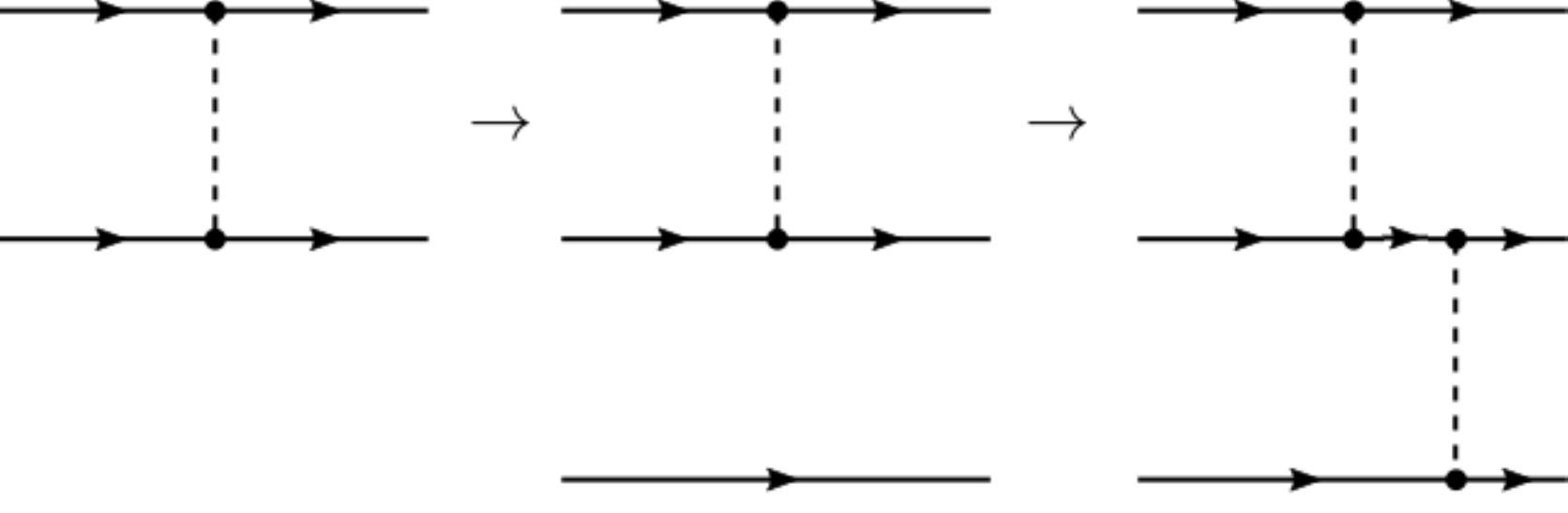}
	\caption{The minimal order possible in a connected Feynman diagram with $N=3$ is $m=2$. The construction is made from the connected diagram with $N=2$ and $m-1$. For arbitrary $N$, the construction is generalizable, we have that the minimal order possible is $m=N-1$.}
	\label{SixLegs}
\end{figure}

\subsection{Sum over all the possible decompositions and general recurrence formula for $\mathbf{\mathcal{N}_{c\,m}^{(N)}}$}

Expression (\ref{Particular}) is only one particular case of (\ref{order}) and (\ref{legs}). If we add all other possible cases, we obtain $\mathfrak{N}_{m}^{(N)}$, namely, the number of total contractions in (\ref{Contract}),

\begin{align}\label{Total}
\mathfrak{N}_{m}^{(N)}=\sum_{(n_{1},\cdots,n_{l}) \in \mathcal{P}_{N}}^{}&\left[\sum_{m_{1}=n_{1}-1}^{m}\cdots\sum_{m_{l}=n_{l}-1}^{m}\sum_{m_{l+1}=0}^{m}\delta_{m_{1}+\cdots+m_{l+1},m}\frac{1}{d_{1}!d_{2}!\cdots d_{r}!}\right.\nonumber\\ &\,\,\,\,\,\,\,\,\,\,\,\,\,\,\,\,\,\,\,\,\,\,\left.\times\frac{(N!)^{2}}{n_{1}!n_{2}!\cdots n_{l}!}\times\frac{m!}{m_{1}!m_{2}!\cdots m_{l+1}!}\mathcal{N}_{c\,m_{1}}^{(n_{1})}\mathcal{N}_{c\,m_{2}}^{(n_{2})}\cdots \mathcal{N}_{c\,m_{l}}^{(n_{l})}\mathfrak{D}_{m_{l+1}}\right].
\end{align}
where $\mathcal{P}_{N}$ is the numerical partition set of $N$. The numbers $\mathfrak{N}_{m}^{(N)}$ and $\mathfrak{D}_{m}$ are given in expressions (\ref{TotalCont}) and (\ref{Vacuum}), respectively. Another way to find this is by using the generating function method mentioned in the introduction. In Appendix \ref{A3} we show that this method leads to expression (\ref{Total}). The index $l$ depends on each partition. The Kronecker delta guarantees that, for one partition of $N$, we have a sum over the weak compositions of $m$, with $N_{i}-1\leq m_{i} \leq m$ for $i \in \{1,\cdots,l\}$, and  $0\leq m_{l+1} \leq m$. (Compositions are partitions where the order of the addends matters. Weak $l$-compositions of a number are all the possible choices $(a_{1},a_{2},\cdots,a_{l})$ such that $a_{i}\ge 0$ and $m=a_{1}+a_{2}+\cdots+a_{l}$. In a $l$-compositions of $m$ we have $a_{i}>0$ \cite{Stanley}.)

From equation (\ref{Total}), it is possible to recursively find the values of $\mathcal{N}_{c\,m}^{(N)}$. (Particularly, it allows a different recurrence for each $N$.) Let us write these recurrences for $N=1,2$ and $3$.

For $N=1$ (two external legs), we have a unique partition 1=1. Therefore, $l=1$ and
\begin{equation}
\mathfrak{N}_{m}^{(1)}=\sum_{m_{1}=0}^{m}\sum_{m_{2}=0}^{m}\delta_{m_{1}+m_{2},m}\frac{m!}{m_{1}!m_{2}!}\mathcal{N}_{c\,m_{1}}^{(1)}\mathfrak{D}_{m_{2}}\label{1}.
\end{equation}	
(See this recurrence in \cite{Castro1}.) 

For $N=2$ (four external legs), the partitions are  $(1+1)$ and $(2)$, with $l=2$ and $l=1$, respectively. So, We have

\begin{align} \label{2}
\mathfrak{N}_{m}^{(2)}=2&\sum_{m_{1}=0}^{m}\sum_{m_{2}=0}^{m}\sum_{m_{3}=0}^{m}\delta_{m_{1}+m_{2}+m_{3},m}\frac{m!}{m_{1}!m_{2}!m_{3}!}\mathcal{N}_{c\,m_{1}}^{(1)}\mathcal{N}_{c\,m_{2}}^{(1)}\mathfrak{D}_{m_{3}}\nonumber \\&+2\sum_{m_{1}=1}^{m}\sum_{m_{2}=0}^{m}\delta_{m_{1}+m_{2},m}\frac{m!}{m_{1}!m_{2}!}\mathcal{N}_{c\,m_{1}}^{(2)}\mathfrak{D}_{m_{2}}.
\end{align}

For $N=3$ (six external legs), the partitions are  $(1+1+1)$, $(2+1)$ and $(3)$, with $l=3$, $l=2$ and $l=1$, respectively. So, We have

 \begin{align} \label{3}
 	\mathfrak{N}_{m}^{(3)}=6\sum_{m_{1}=0}^{m}\sum_{m_{2}=0}^{m}&\sum_{m_{3}=0}^{m}\sum_{m_{4}=0}^{m}\delta_{m_{1}+m_{2}+m_{3}+m_{4},m}\frac{m!}{m_{1}!m_{2}!m_{3}!m_{4}!}\mathcal{N}_{c\,m_{1}}^{(1)}\mathcal{N}_{c\, m_{2}}^{(1)}\mathcal{N}_{c\,m_{3}}^{(1)}\mathfrak{D}_{m_{4}}\nonumber \\ +18&\sum_{m_{1}=1}^{m}\sum_{m_{2}=0}^{m}\sum_{m_{3}=0}^{m}\delta_{m_{1}+m_{2}+m_{3},m}\frac{m!}{m_{1}!m_{2}!m_{3}!}\mathcal{N}_{c\,m_{1}}^{(2)}\mathcal{N}_{c\,m_{2}}^{(1)}\mathfrak{D}_{m_{3}}\nonumber \\&+6\sum_{m_{1}=2}^{m}\sum_{m_{2}=0}^{m}\delta_{m_{1}+m_{2},m}\frac{m!}{m_{1}!m_{2}!}\mathcal{N}_{c\, m_{1}}^{(3)}\mathfrak{D}_{m_{2}}.
 \end{align}

Recurrence (\ref{1}) determines the numbers $\mathcal{N}_{c\,m}^{(1)}$, which can be used in the recurrence (\ref{2}) to find the numbers $\mathcal{N}_{c\,m}^{(2)}$, and so on.

\section{Recurrence simplification}\label{R}

\begin{figure}
	\centering
		\includegraphics[width=0.5\textwidth]{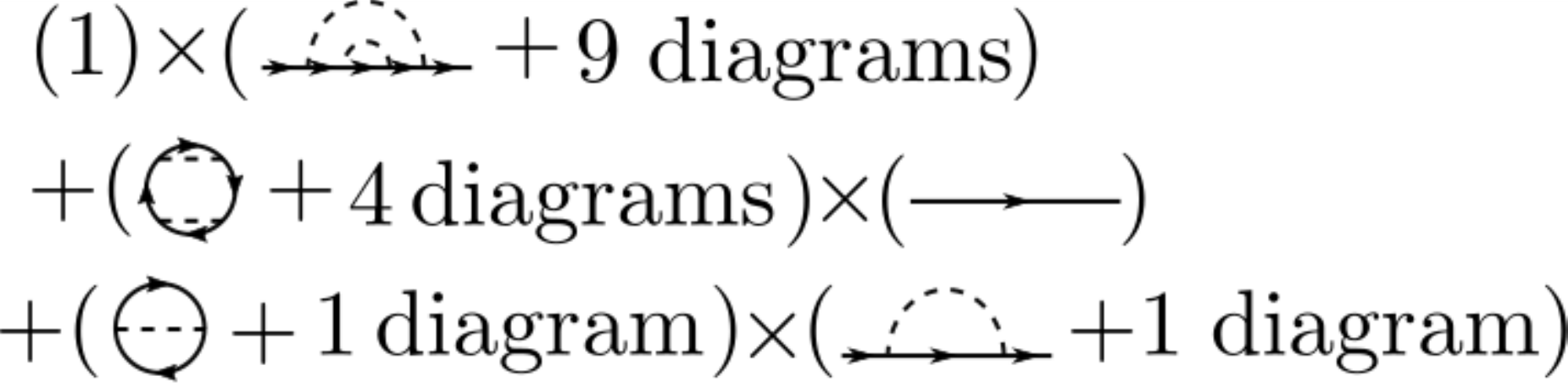}
	\caption{A diagrammatic form to create the nineteen Feynman diagrams with two external legs and $m=2$. This picture suggests the formula $\mathfrak{D}_{0}\mathcal{N}_{c\,2}^{(1)}+\mathfrak{D}_{1}\mathcal{N}_{c\,1}^{(1)}+\mathfrak{D}_{2}\mathcal{N}_{c\,0}^{(1)}$  for the contractions that generate these diagrams. The correct formula (\ref{1}) is weighed by a binomial coefficient.}
	\label{Care}
\end{figure}

Recurrence (\ref{Total}) has an uncomplicated interpretation in terms of a discrete convolution in the number of contractions that generates the arbitrary component associated with the perturbative order of  each component. This was correctly noticed in the recurrences obtained in Ref.\cite{Kugler}. Also, some care must be taken when converting diagramatic expressions like (\ref{graph}) into numerical convolutions, because, as in (\ref{graph}), combinatorial weights can be involved in the discrete convolution, see fig \ref{Care}.

In our case, the terms of the discrete convolutions are indexed by weak compositions \cite{Stanley}. For a precise recurrence computation, this can be a problem, since this would require evaluating a huge number of possibilities. Fortunately, expression (\ref{Total}) can be greatly simplified. In particular, we have    

\begin{equation}\label{Simple}
\mathfrak{N}_{m}^{(N)}=N\sum_{j=0}^{m}\binom{m}{j}\mathcal{N}_{c\,j}^{(1)}\mathfrak{N}_{m-j}^{(N-1)}+ N\sum_{i=2}^{N}\frac{(N-1)^{2}(N-2)^{2}\cdots(N-i+1)^{2}}{(i-1)!}\sum_{j=i-1}^{m}\binom{m}{j}\mathcal{N}_{c\,j}^{(i)}\mathfrak{N}_{m-j}^{(N-i)},
\end{equation}
where $\mathfrak{N}_{l}^{(0)}=\mathfrak{D}_{l}=(2l)!$.

\subsection{Proof of the simplified recurrence formula}

To prove this from expression (\ref{Total}), note that in (\ref{Total}) we have a sum indexed over the possible partitions of $N$. Consider the set of all the partitions $\mathcal{A}_{n_{i}} \subset \mathcal{P}_{N}$ that contain the arbitrary number $n_{i}$. Since $N=n_{i}+(N-n_{i})$, the number of such partitions is identical to the total number of partitions in $\mathcal{P}_{N-n_{i}}$. (That is to say, there exists a bijection between $\mathcal{A}_{n_{i}}$ and $\mathcal{P}_{N-n_{i}}$.) Considering an arbitrary partition of $N$ that contain $n_{i}$, the redefinition $n_{i}\to n_{a}$ and $n_{j+1} \to n_{j}$ for $i\leq j \leq l$, the corresponding term in (\ref{Total}) can be written as

  \begin{align}\label{P1}
  	\sum_{m_{a}=n_{a}-1}^{m}\frac{m!}{m_{a}!(m-m_{a})!}\mathcal{N}_{c\,m_{a}}^{(n_{a})}&\left[\sum_{m_{1}=n_{1}-1}^{m-m_{a}}\cdots\sum_{m_{l-1}=n_{l-1}-1}^{m-m_{a}}\sum_{m_{l}=0}^{m-m_{a}}\delta_{m_{1}+\cdots+m_{l},m-m_{a}}\frac{1}{d_{1}!d_{2}!\cdots d_{r}!}\right.\nonumber\\ &\,\,\,\,\,\,\,\,\,\,\,\,\,\,\,\,\,\,\,\,\,\,\left.\times\frac{(N!)^{2}}{(n_{1}!n_{2}!\cdots n_{l-1}!)n_{a}!}\times\frac{(m-m_{a})!}{m_{1}!m_{2}!\cdots m_{l}!}\mathcal{N}_{c\,m_{1}}^{(n_{1})}\mathcal{N}_{c\,m_{2}}^{(n_{2})}\cdots \mathcal{N}_{c\,m_{l-1}}^{(n_{l-1})}\mathfrak{D}_{m_{l}}\right],
  \end{align}
where, in the new definition, the Kronecker delta guarantee that $m_{1}+\cdots+m_{l}=m-m_{a}$. Using $N=d_{1}n_{1}+\cdots+d_{r}n_{r}$, we have

\begin{equation}\label{P2}
\frac{1}{d_{1}!d_{2}!\cdots d_{r}!}\times\frac{(N!)^{2}}{(n_{1}!n_{2}!\cdots n_{l-1}!)n_{a}!}=\frac{1}{d_{1}!d_{2}!\cdots d_{r}!}\times\frac{N\left[(N-1)!\right]^{2}}{(n_{1}!n_{2}!\cdots n_{l-1}!)n_{a}!}\times\left(d_{1}n_{1}+\cdots+d_{r}n_{r}\right).
\end{equation}

From the previous equation, it is obvious that we have $r$ different choices for the index $n_{a}$ associated with the partition related to (\ref{P1}). Therefore, we can decompose (\ref{P1}) in $r$ terms. The term associated with $n_{a}$ has the weight

\begin{align}
\frac{1}{d_{1}!\cdots d_{a}!\cdots d_{r}!}\times\frac{N\left[(N-1)!\right]^{2}}{(n_{1}!n_{2}!\cdots n_{l-1}!)n_{a}!}\times d_{a}n_{a}&=\frac{N(N-1)^{2}\cdots(N-n_{a}+1)^{2}}{(n_{a}-1)!}\nonumber\\ &\,\,\,\,\,\times\left[\frac{1}{d_{1}!\cdots (d_{a}-1)!\cdots d_{r}!}\times\frac{\left[(N-n_{a})!\right]^{2}}{n_{1}!n_{2}!\cdots n_{l-1}!}\right].
\end{align}

So, according to expression (\ref{P1}) we have $r$ terms associated in the following format

\begin{align}\label{P3}
	\frac{N(N-1)^{2}\cdots(N-n_{a}+1)^{2}}{(n_{a}-1)!}\sum_{m_{a}=n_{a}-1}^{m}\binom{m}{m_{a}}\mathcal{N}_{c\,m_{a}}^{(n_{a})}&\left[\sum_{m_{1}=n_{1}-1}^{m-m_{a}}\cdots\sum_{m_{l-1}=n_{l-1}-1}^{m-m_{a}}\sum_{m_{l}=0}^{m-m_{a}}\delta_{m_{1}+\cdots+m_{l},m-m_{a}}\right.\nonumber \\ &\left.\;\;\;\;\;\;\;\;\times\frac{1}{d_{1}!\cdots (d_{a}-1)!\cdots d_{r}!}\times\frac{\left[(N-n_{a})!\right]^{2}}{(n_{1}!n_{2}!\cdots n_{l-1}!)}\right.\nonumber\\ &\left.\,\,\,\,\;\;\;\;\;\;\times\frac{(m-m_{a})!}{m_{1}!m_{2}!\cdots m_{l}!}\times \mathcal{N}_{c\,m_{1}}^{(n_{1})}\mathcal{N}_{c\,m_{2}}^{(n_{2})}\cdots \mathcal{N}_{c\, m_{l-1}}^{(n_{l-1})}\mathfrak{D}_{m_{l}}\right].
\end{align}

The important fact about these $r$ terms associated with the initial arbitrary partition of $N$ is that all of them have the same form, and we can consider (\ref{P3}) as a generic term. The factor in the square bracket of (\ref{P3}) corresponds to a partition of $N-n_{a}$ and is identical to the associated term in $\mathfrak{N}_{m-m_{a}}^{(N-n_{a})}$. (See expression (\ref{Total}).) From the bijection $\mathcal{A}_{n_{a}}\longleftrightarrow \mathcal{P}_{N-n_{a}}$, we exhaust all the possibilities for the other partitions of $N$ that contain the element $n_{a}$, getting all the partitions in $\mathcal{P}_{N-n_{a}}$ and therefore generating all the terms of $\mathfrak{N}_{m-m_{a}}^{(N-n_{a})}$: 

\begin{equation}
\frac{N(N-1)^{2}\cdots(N-n_{a}+1)^{2}}{(n_{a}-1)!}\sum_{m_{a}=n_{a}-1}^{m}\binom{m}{m_{a}}\mathcal{N}_{c\,m_{a}}^{(n_{a})}\mathfrak{N}_{m-m_{a}}^{(N-n_{a})}.
\end{equation}

For all the other possible values of $n_{a}$, we repeat the process. Since (\ref{Total}) is associated with all the partitions of $N$, the decomposition (\ref{P2}) and the bijection $\mathcal{A}_{n_{a}}\longleftrightarrow \mathcal{P}_{N-n_{a}}$ guarantee the validity of relation (\ref{Simple}). For $n_{a}=1$, it is clear that the factor $(N-1)^{2}\cdots(N-n_{a}+1)^{2}/(n_{a}-1)!$ does not appear.

\subsection{The number of $\mathbf{m}$-order connected Feynman diagrams $\mathbf{h_{m}^{(N)}}$} 

\begin{figure}[H]
	\centering
		\includegraphics[width=0.7\textwidth]{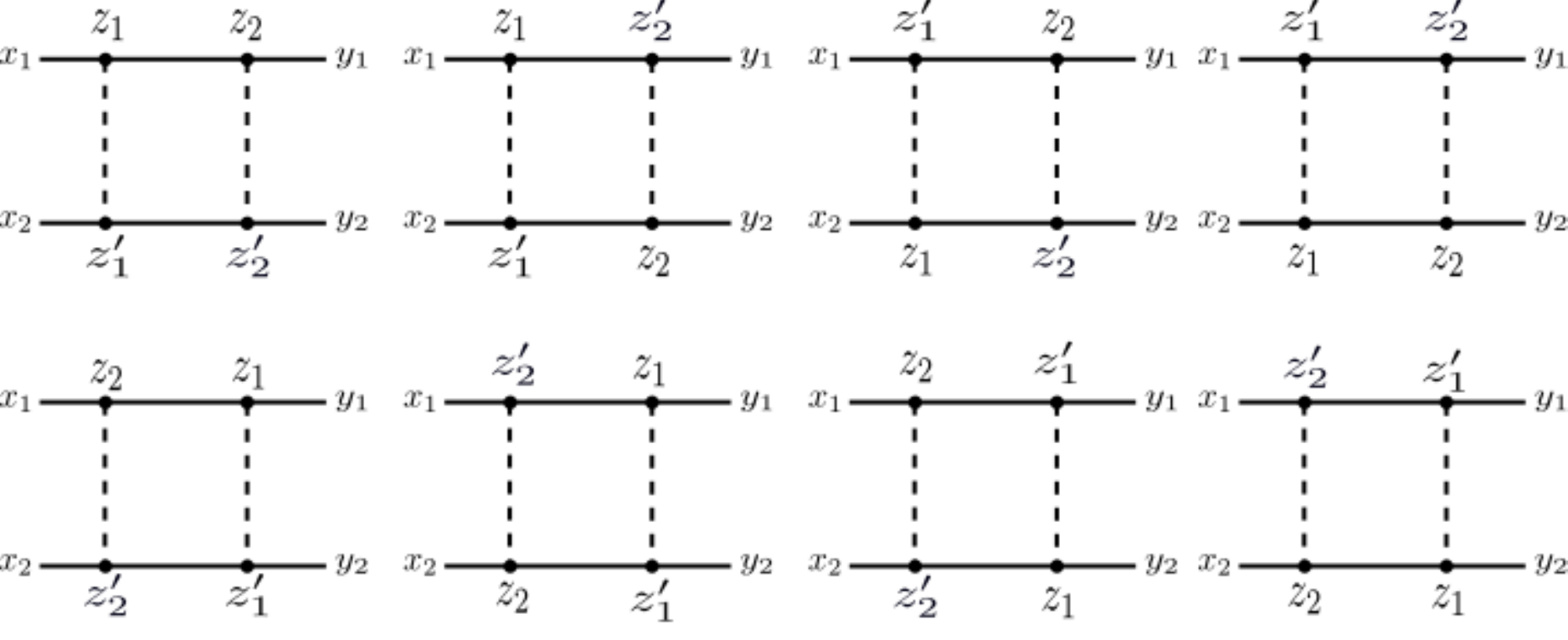}
	\caption{The $2^{2}\times 2!=8$ contractions that lead to the same connected Feynman diagram. In order to avoid counting these equivalent contractions in $\mathfrak{N}_{c\,m}^{(N)}$, we simply divide by the factor $2^{m}m!$.}
	\label{Contra}
\end{figure}

Now, remember that the number $\mathcal{N}_{c\,m}^{(N)}$ is the total number of contractions that generate $m$-order connected Feynman diagrams with $2N$ {\it fixed} external legs. Some of these contractions generate the same Feynman diagram. In particular, every $m$-order connected Feynman diagram with $N>0$, has multiplicity (i.e., different equivalent contractions) equal to $2^{m}m!$ or the same symmetry factor. To see this, note that every dashed line can be chosen from $(2m)!!$ ways. (The first dashed line can be chosen from the $m^{\prime}$s $U(z_{i},z_{i}^{\prime})$ and every $U(z_{i},z_{i}^{\prime})$ in two ways. Therefore, we have $2m$ possibilities. For the second dashed line, we have $(2m-2)$ possibilities and so on. See the example in Fig.\ref{Contra}). Therefore, the number of different $m$-order connected Feynman diagrams with $2N$ external legs $h_{m}^{(N)}$ is

\begin{equation}\label{h1}
h_{m}^{(N)}=\frac{\mathcal{N}_{c\,m}^{(N)}}{2^{m}m!}.
\end{equation}

 Table \ref{Table1} shows the initial series of values for $h_{m}^{(N)}$.

	\begin{center}
	\captionof{table}{Initial series of values for $h_{m}^{(N)}$.}
	\label{Table1}
	\begin{tabular}{| c | c |c |c | c | c |c |c |}
		\hline
		& $h_{m}^{(1)}$ & $h_{m}^{(2)}$ & $h_{m}^{(3)}$ & $h_{m}^{(4)}$  & $h_{m}^{(5)}$  & $h_{m}^{(6)}$  & $h_{m}^{(7)}$ \\ \hline $m=0$ & $1$ & $0$ & $0$ & $0$   & $0$ & $0$  & $0$    \\ \hline
		$m=1$ & $2$ & $1$ & $0$ & $0$   & $0$ & $0$  & $0$    \\ \hline
		$m=2$ & $10$ & $13$ & $6$ & $0$   & $0$ & $0$  & $0$   \\ \hline
		$m=3$ & $74$ & $165$ & $172$ & $72$   & $0$ & $0$  & $0$  \\ \hline
		$m=4$ & $706$ & $2273$ & $3834$ & $3438$   & $1320$ & $0$  & $0$  \\ \hline
		$m=5$ & $8162$ & $34577$ & $81720$ & $115008$   & $91968$ & $32760$  & $0$  \\ \hline
		$m=6$ & $110410$ & $581133$ & $1775198$ & $3432864$   & $4227840$ & $3082080$  & $1028160$  \\ \hline
		$m=7$ & $1708394$ & $10749877$ & $40320516$ & $99431808$   & $166020720$ & $184019040$  & $124126560$  \\ \hline
	\end{tabular}
\end{center}

\begin{figure}[H]
	\centering
		\includegraphics[width=0.7\textwidth]{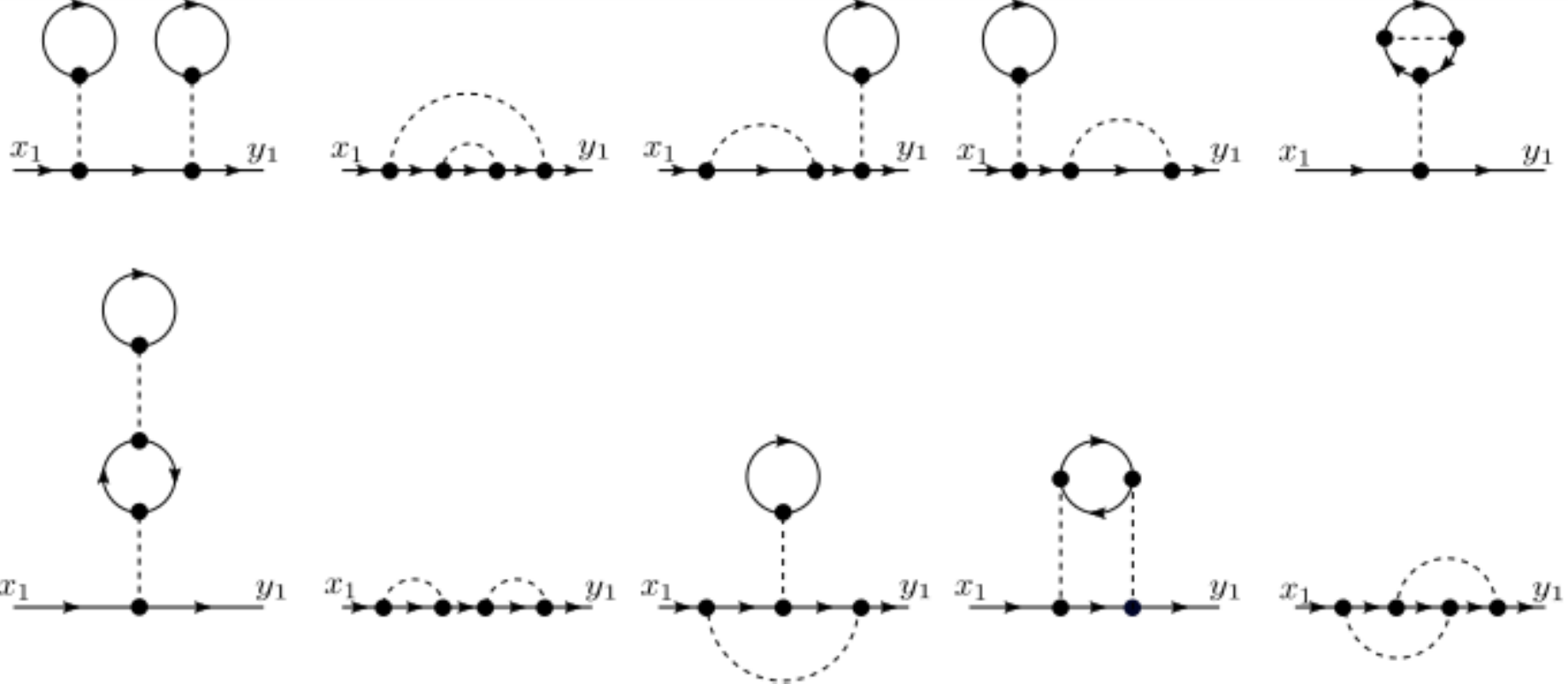}
	\caption{The ten connected Feynman diagrams for $m=2$ and $N=1$.}
	\label{First10}
\end{figure}

\begin{figure}[H]
	\centering
		\includegraphics[width=0.6\textwidth]{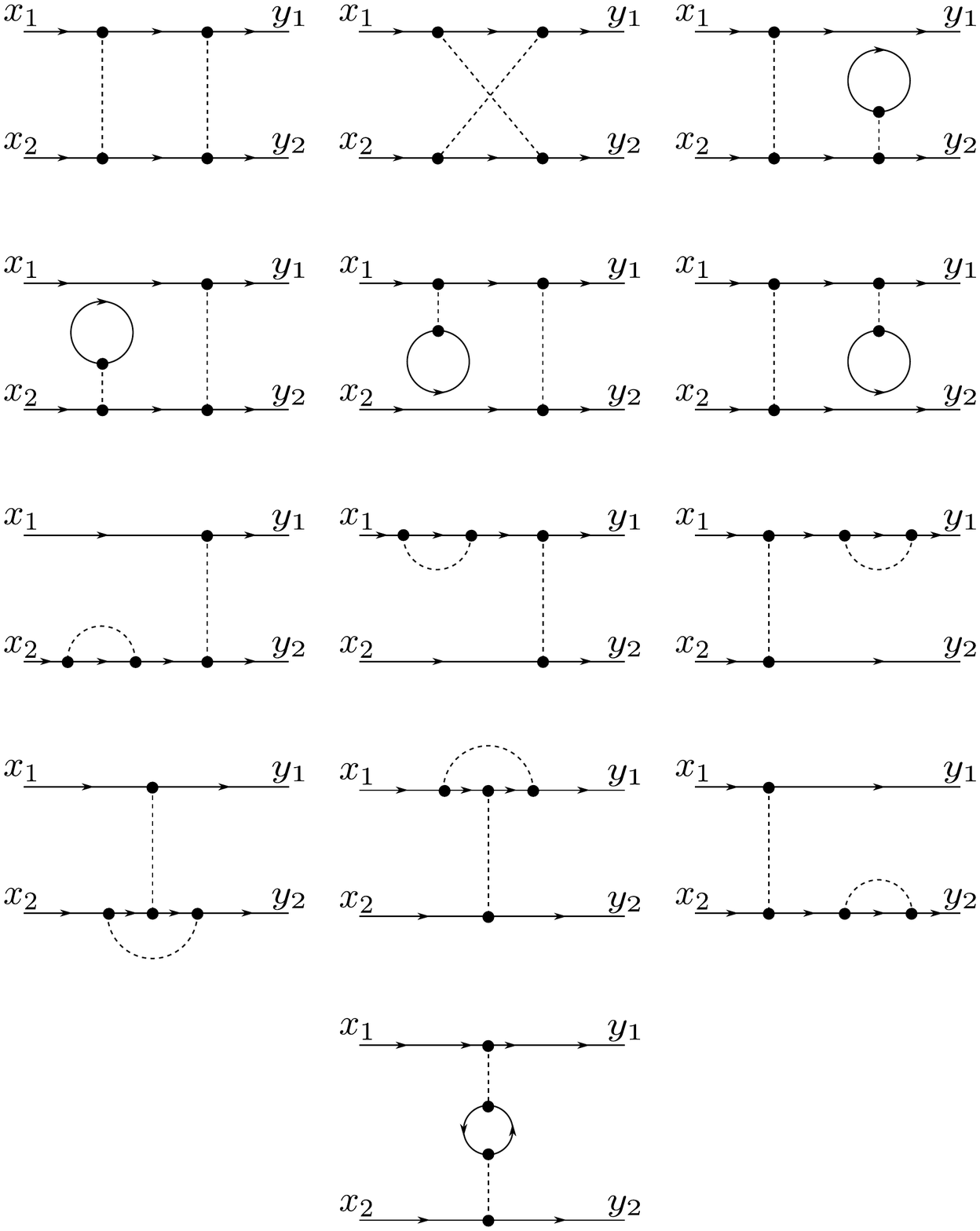}
	\caption{The thirteen connected Feynman diagrams for $m=2$ and $N=2$. Note that the external legs are labeled. For unlabeled external legs, we have only 8 connected Feynman diagrams.}
	\label{13D}
\end{figure}

\begin{figure}[H]
	\centering
		\includegraphics[width=0.7\textwidth]{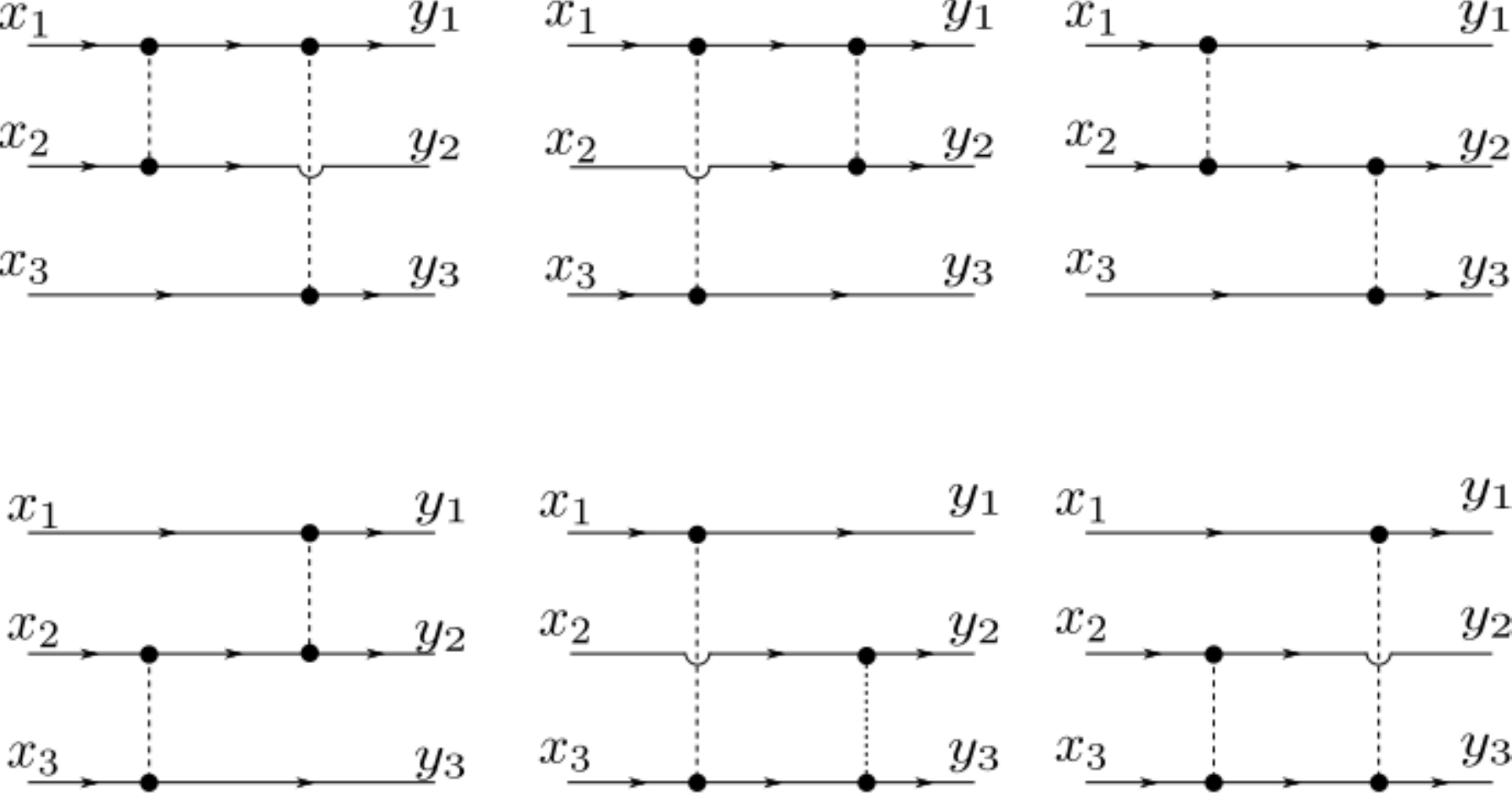}
	\caption{The six connected Feynman diagrams for $m=2$ and $N=3$. Note that the external legs are labeled. For unlabeled external legs, all these diagrams would be equivalent.}
	\label{Second6}
\end{figure}

The sequence $h_{m}^{(1)}$ corresponds to the OEIS sequence $A000698$. The same values for $h_{m}^{(2)}$ and $h_{m}^{(3)}$ are given as long as $m \le 6$ in formulas (25) and (29) of Ref.\cite{Gopala2}. In figures \ref{First10}, \ref{13D} and \ref{Second6}, we show the connected diagrams corresponding to $h_{2}^{(1)}$, $h_{2}^{(2)}$ and $h_{2}^{(3)}$, see directly the thirteen diagrams of Fig. \ref{13D} in ref.\cite{Bachmann} used for M{\o}ller and Bhabba scattering. Note that the six diagrams for $m=2$ and $N=3$ are considered different since the external legs are labeled. If the external legs were not labeled, the counting would be different, in the case $\{m=2,N=2\}$, we would have eight different diagrams, and for $\{m=2,N=3\}$, we would have only one. For unlabeled external, legs the counting is very different and, in principle, more difficult.  We will continue considering only Feynman diagrams with labeled external legs. 

The numerical solutions of recurrences (\ref{Simple}) are constructive, that is to say, they are solved by beginning with case $N=1$ until finite order $m$. Then, all these values are used to solve case $N=2$ until finite order $m$, and so on. For example, using the program MATHEMATICA, we have calculated the exact values of $\mathcal{N}_{c\,m}^{(N)}$ up to $m=3000$ for the cases $N=1,2,\cdots,7$ in a few minutes.

\section{Exact solution for cases $N=1$ and $N=2$, asymptotic expansion} \label{E}

Recurrences (\ref{Simple}) can be solved exactly for cases $N=1$ and $N=2$. This allows the calculation of many terms in the asymptotic expansion ($m\to \infty$) of $h_{m}^{(N)}$. 

\subsection{Exact solution for the case $N=1$}

In Ref.\cite{Castro1}, an explicit formula for $\mathcal{N}_{c\,m}^{(1)}$ is obtained:

\begin{equation}
\mathcal{N}_{c\,m}^{(1)}=\sum_{n=1}^{m}\mathcal{C}_{n}^{m}\left(\mathfrak{N}_{n}^{(1)}-\mathfrak{D}_{n}\right)\label{firstCountFormula},
\end{equation}
with
\begin{equation}
\mathcal{C}_{n}^{m}=\sum_{i=1}^{m-n}(-1)^{i}\sum_{a_{1},\cdots,a_{i}=1}^{\infty}\delta_{a_{1}+\cdots+a_{i},m-n}\binom{m}{m-a_{1}}\binom{m-a_{1}}{m-a_{1}-a_{2}}\cdots\binom{m-a_{1}-\cdots-a_{i-1}}{m-a_{1}-\cdots-a_{i-1}-a_{i}}\prod_{j=1}^{i}\mathfrak{D}_{a_{j}},\label{Combin}
\end{equation}
for $n<m$, and $\mathcal{C}_{m}^{m}=1$ for $m \in \mathbb{N}$.  The above formula can be written as

\begin{equation}
\mathcal{C}_{n}^{m}=-\frac{m!}{n!}\sum_{i=0}^{m-n-1}(-1)^{i}\sum_{a_{1},\cdots,a_{i+1}=1}^{\infty}\delta_{a_{1}+\cdots+a_{i+1},m-n}\prod_{j=1}^{i+1}\frac{(2a_{j})!}{a_{j}!},\label{Combin2}.
\end{equation}
If we compare the expression (\ref{Combin2}) to the Arqu\`es-Walsh sequence formula \cite{Castro1}, we obtain for $n<m$

\begin{equation}\label{Simplified}
\mathcal{C}_{n}^{m}=-2(m-n)\binom{m}{n}\mathcal{N}_{c\,m-n-1}^{(1)}.
\end{equation}

Since $\mathcal{N}_{c\,l}^{(1)}$ is always positive $\forall \,l$, the symbols $\mathcal{C}_{n}^{m}$ are negative for $n<m$. In particular, for expression (\ref{firstCountFormula}), we have

\begin{equation}
\mathfrak{N}_{m}^{(1)}-\mathfrak{D}_{m}>-\sum_{n=1}^{m-1}\mathcal{C}_{n}^{m}\left(\mathfrak{N}_{n}^{(1)}-\mathfrak{D}_{n}\right)>0
\end{equation}

 For arbitrary and finite $m$, the formula (\ref{Simplified}) simplifies the calculation of the symbols $\mathcal{C}_{n}^{m}$ to the case $n \lesssim m$. (For example, if $n=m-5$, in formula \ref{Simplified} we only need to know $\mathcal{N}_{c\,m-(m-5)-1}=\mathcal{N}_{c\,4}$, which can be easily calculated from the recurrences.) In this case, it is only necessary to know the first values of $\mathcal{N}_{c\,k}^{(1)}$, which are obtained iterating (\ref{Simple}) for $N=1$. In particular,

\begin{equation}
\mathcal{C}_{m-1}^{m}=-2m
\end{equation}
\begin{equation}
\mathcal{C}_{m-2}^{m}=-8m(m-1)
\end{equation}
\begin{equation}
\mathcal{C}_{m-3}^{m}=-80m(m-1)(m-2)
\end{equation}
\begin{equation}
\mathcal{C}_{m-4}^{m}=-1184m(m-1)(m-2)(m-3)
\end{equation}
\begin{equation}
\mathcal{C}_{m-5}^{m}=-22592m(m-1)(m-2)(m-3)(m-4)
\end{equation}
\begin{equation}
\mathcal{C}_{m-6}^{m}=-522368m(m-1)(m-2)(m-3)(m-4)(m-5)
\end{equation}
\begin{equation}\label{ck}
\mathcal{C}_{m-k}^{m}=-\frac{2k}{k!} N_{c\, k-1}m(m-1)(m-2)(m-3)(m-4)\cdots(m-k+1).
\end{equation}
\begin{center}
	\captionof{table}{Initial values for $\vert\mathcal{C}_{n}^{m}\vert$.}
\label{Table2}
	\begin{tabular}{| c | c |c |c | c | c |c|}
		\hline
		& $n=1$ & $n=2$ & $n=3$ & $n=4$  & $n=5$ 
\\ \hline $m=2$ & $4$ & $1$ & $0$ & $0$   & $0$ \\ \hline
		$m=3$ & $48$ & $6$ & $1$ & $0$     & $0$ \\ \hline
		$m=4$ & $ 1920$ & $96$ & $8$ & $1$     & $0$ \\ \hline
		$m=5$ & $ 142080$ & $ 4800$ & $ 160$ & $10$  & $1$  \\ \hline
		$m=6$ & $ 16266240$ & $ 426240$ & $ 9600$ & $ 240$ & $12$   \\ \hline
		
	\end{tabular}
\end{center}

\subsection{Asymptotic expansion for the case $N=1$}

We are interested in an asymptotic expansion for $h_{m}^{(1)}$, when $m\to \infty$. Using $\mathfrak{N}_{m}^{(1)}-\mathfrak{D}_{m}=(2m)(2m)!$, the number of different connected Feynman diagrams with two external legs from (\ref{firstCountFormula}) is

\begin{equation}\label{h}
h_{m}^{(1)}=\frac{\mathcal{N}_{c\,m}^{(1)}}{2^{m}m!}=\frac{\mathfrak{N}_{m}^{(1)}-\mathfrak{D}_{m}}{2^{m}m!}\left[1+\sum_{n=1}^{m-1}\mathcal{C}_{n}^{m}\frac{\mathfrak{N}_{n}^{(1)}-\mathfrak{D}_{ n}}{\mathfrak{N}_{m}^{(1)}-\mathfrak{D}_{m}}\right]=\frac{m!m}{2^{m-1}}\binom{2m}{m}\left[1+\sum_{k=1}^{m-1}\frac{(m-k)(2[m-k])!}{m(2m)!}\mathcal{C}_{m-k}^{m}\right].
\end{equation}
In the last step, we use $k=m-n$.

\subsubsection{First contribution to the asymptotic expansion (case $k\ll m$)}

 Now, let's focus on the square bracket term in (\ref{h}). It is easy to notice, by using (\ref{Combin2}), that each term in the sum, is a quotient of polynomials in $m$. To see this, choose a fixed value for $k$ and use (\ref{ck}). The terms in question are proportional to

\begin{equation}
\frac{(m-k)}{(2m)(2m-1)(2m-3)(2m-5)\cdots(2m-(2k-1))}
\end{equation}
for $k<m$ and $k\in \mathbb{N}$. By adding the first $\ell$ terms, it is not hard to see that

\begin{equation}\label{q}
P(m,\ell)=\sum_{k=1}^{\ell}\frac{(m-k)(2[m-k])!}{m(2m)!}\mathcal{C}_{m-k}^{m}=-\frac{A^{(\ell)}m^{\ell}+B^{(\ell)}m^{\ell-1}+\cdots+X^{(\ell)}m^{2}+Y^{(\ell)}m+Z^{(\ell)}}{(2m)(2m-1)(2m-3)(2m-5)\cdots(2m-(2\ell-1))}
\end{equation}
where the numbers $A^{(\ell)},B^{(\ell)},\cdots, Z^{(\ell)}$ are integers generated by the usual algebraic operations when we factor the terms in the sum.

Let's see the cases $\ell=1,2$ and $3$. For $\ell=1$,

\begin{equation}
P(m,1)=\sum_{k=1}^{1}\frac{(m-k)(2[m-k])!}{m(2m)!}\mathcal{C}_{m-k}^{m}=\frac{(m-1)[2(m-1)]!}{m(2m)!}\mathcal{C}_{m-1}^{m}=-\frac{2m-2}{(2m)(2m-1)}
\end{equation}

for $\ell=2$

\begin{equation}
P(m,2)=\sum_{k=1}^{2}\frac{(m-k)(2[m-k])!}{m(2m)!}\mathcal{C}_{m-k}^{m}=-\frac{2m-2}{(2m)(2m-1)}-\frac{4m-8}{(2m)(2m-1)(2m-3)}=-\frac{4m^{2}-6m-2}{(2m)(2m-1)(2m-3)}
\end{equation}

for $\ell=3$

\begin{equation}
P(m,3)=\sum_{k=1}^{3}\frac{(m-k)(2[m-k])!}{m(2m)!}\mathcal{C}_{m-k}^{m}=-\frac{8m^{3}-28m^{2}+36m-44}{(2m)(2m-1)(2m-3)(2m-5)}
\end{equation}

and so on.

The quotient of polynomials (\ref{q}) has uncomplicated analytical properties. For example, it has at least $\ell+1$ poles of order 1. What is important in this case is that the point in the infinity is regular, and the complex form of (\ref{q}) is analytical in some neighborhood of $m=\infty$. Therefore, $P(m,\ell)$ admits an analytical Taylor expansion in $m=\infty$. By making the transformation $w=1/m$, we see that the convergence radius of the Taylor expansion in $\omega=0$ is $2/(2\ell-1)$ (with the assumption that, for all $\ell$, $z_{0}=\ell-1/2$ is a pole). For increasing values of $\ell$, more and more poles appear in the real positive axis and the convergence radius of the Taylor series in $\omega=0$ tends to zero.

Fortunately, this does not prevent the asymptotic analysis in $m\to \infty$. Note that, for fixed $\ell$, only the first $a$ left-hand-side terms of (\ref{q}) contribute in the first $a$ Taylor-series terms of the right hand side, when $m>\ell-1/2$. This last condition guarantees that $m$ is inside the convergence radius, and  the Taylor expansion of the left-hand-side terms can be added term by term. In particular the left hand side terms of (\ref{q}) have the next Taylor expansion in the infinity

\begin{equation}\label{Taylor}
\frac{2k}{2^{k-1}k!} \mathcal{N}_{c\, k-1}\frac{(m-k)}{(2m)(2m-1)(2m-3)(2m-5)\cdots(2m-(2k-1))}=\frac{a_{1}}{m^{k}}+\frac{a_{2}}{m^{k+1}}+\cdots,\,\,\,\,\,\,\,\, m>k-\frac{1}{2}.
\end{equation}

From this expression, it becomes clear that we only need to sum up the $a$ initial Taylor terms of the $a$ initial left-hand-side terms of (\ref{q}) to get the first $a$ Taylor-series terms of the right-hand-side. For $k>a$, the Taylor terms of (\ref{Taylor}) are of order $O(1/m^{a+n})$, with $n \in \mathbb{N}$, and do not contribute to the first $a$ Taylor terms of the right-hand-side of (\ref{q}). This analysis is valid for $m$ arbitrarily large and {\it finite}, and we can interpret the Taylor-series in the infinity with convergent radius zero as the asymptotic expansion of $h_{m}^{(1)}$.

\subsubsection{Second contribution to the asymptotic expansion (case $k\le m$}

 In the expansion (\ref{h}) for arbitrarily large $m$, we only analize the contribution of the terms $k\ll m$. We now show that the terms $k\lesssim m$ are also present in the contribution to the first asymptotic terms of the expansion. Using (\ref{Combin2}), we have the following term, for $k=m-1$

\begin{equation}
\frac{(m-1)!2!}{0!(2m!)}\sum_{i=0}^{m-2}(-1)^{i+1}\sum_{a_{1},\cdots,a_{i+1}=1}^{m-1}\delta_{a_{1}+\cdots+a_{i+1},m-1}\prod_{j=1}^{i+1}\frac{(2a_{j})!}{a_{j}!}\label{caso1}.
\end{equation}
For a fixed $i$, we see that the term of (\ref{caso1}) represent compositions of $m-1$ with $i+1$ elements. (The elements are the $a_{k}$ coefficients.) The first terms are

\begin{align}\label{Expan}
-\frac{(m-1)!2!}{0!(2m!)}&\left[\frac{[2(m-1)]!}{(m-1)!}-2\frac{[2(m-2)]!}{(m-2)!}\frac{2!}{1!}-2\frac{[2(m-3)]!}{(m-3)!}\frac{4!}{2!}+\cdots\right. \nonumber \\ & +3\frac{[2(m-3)]!2!2!}{(m-3)!1!1!}+6\frac{[2(m-4)]!4!2!}{(m-4)!2!1!}+3\frac{[2(m-5)]!4!4!}{(m-5)!2!2!}\cdots \nonumber \\ & \left.-4\frac{[2(m-4)]!2!2!2!}{(m-4)!1!1!1!}-12\frac{[2(m-5)]!4!2!2!}{(m-5)!2!1!1!}-\cdots\right].
\end{align}
For example, the second term in the second line represents the composition $m-1=(m-4)+2+1$, and the multiplicative factor 6 represents the possible permutations of these three coefficients. The same analysis is, then, performed for $k=m-2$, $k=m-3$, etc. By a similar analysis, performed beforehand, of equation (\ref{Taylor}), only a finite number of these terms contribute to the first $a$ asymptotic terms of the entire expression (\ref{h}). In Appendix \ref{A2}, we explicitly write all the terms that contribute to $a=6$.

\subsubsection{The total contribution to the conventional asymptotic expansion}

The expansion until $O(1/m^{6})$, considering both contributions, is

\begin{equation}
\left[1+\sum_{k=1}^{m-1}\frac{(m-k)(2[m-k])!}{m(2m)!}\mathcal{C}_{m-k}^{m}\right]\sim 1-\frac{1}{2m}-\frac{3}{4m^{2}}-\frac{19}{8m^{3}}-\frac{191}{16m^{4}}-\frac{2551}{32m^{5}}-\frac{41935}{64m^{6}}\cdots\label{Asym}
\end{equation}
 and the expansion of the central binomial coefficient until order six \cite{Neven} is
\begin{equation}\label{CenBin}
\binom{2m}{m}\sim \frac{4^{m}}{\sqrt{m\pi}}\left[1-\frac{1}{8m}+\frac{1}{128m^{2}}+\frac{5}{1024m^{3}}-\frac{21}{32768m^{4}}-\frac{399}{262144m^{5}}+\frac{869}{4194304m^{6}}\cdots\right].
\end{equation}
By using both these expansions in (\ref{h}), we obtain the expansion for $h_{m}^{(1)}$:

\begin{equation}\label{hm1}
h_{m}^{(1)}\sim \frac{2}{\sqrt{\pi}}m!m^{\frac{1}{2}}2^{m}\left[1-\frac{5}{8m}-\frac{87}{128m^{2}}-\frac{2335}{1024m^{3}}-\frac{381733}{32768m^{4}}-\frac{20512763}{262144m^{5}}-\frac{2706890307}{4194304m^{6}}\cdots\right].
\end{equation}
The first two coefficents match with the ones given in Ref.\cite{Cvitanovic}, which use the functional approach. (In this reference, the variable used is $m^{\prime}=2m$. The prefactor $1/\pi$ in this reference should be $1/\sqrt{\pi}$, this was also noticed also by \citep{Kugler}.) For $m=1000,2000$ and $3000$, our asymptotics match with the exact value in the first 17, 19 and 20 digits, respectively.

\subsection{A new asymptotic contribution: the centered multinomial contribution}

The excellent numerical matching suggests that we have obtained the first six terms ($a \leq$ 6) of the asymptotic expansion. The other terms ($a \geq 6$) can be found by the same process. (The calculation for growing $a$ gets more complicated, since it involves more and more terms of (\ref{h}) and (\ref{Combin2}).) In particular, the terms of (\ref{h}) for $k \lesssim m$ can be written as quotients of multinomial coefficients. For example, consider the last term of (\ref{Expan}),

\begin{equation}
12\frac{(m-1)!\left[2(m-5)\right]!2!4!2!2!}{(2m)!(m-5!)2!1!1!0!}=12\frac{\binom{m-1}{m-5,2,1,1}}{\binom{2m}{2m-10,4,2,2,2}}.
\end{equation}
 All the terms in (\ref{Expan}), and also for the cases $k=m-2,m-3,\cdots$, can be written as quotients of multinomial coefficients. In the limit $k\lesssim m$ studied before, we only consider {\it non-centered} elements of the corresponding multinomial distributions. Every one of these terms is also represented by a numerical partition, which have a unique element depending on $m$.
 
  Other cases not analyzed correspond to centered elements of the multinomial distribution or to numerical partitions with more than one coefficient depending on $m$. Do these terms present asymptotic contribution in $m\to \infty$? To analyze this, let us assume that $m$ is even. A specific example in the case $k=m-1$ would be
 
    \begin{equation}\label{Asy}
    12\frac{(m-1)!(m-6)!2!m!2!2!}{(2m)!(m/2-3!)(m/2)!1!1!0!}=12\frac{\binom{m-1}{m/2-3,m/2,1,1}}{\binom{2m}{m-6,m,2,2,2}}.
    \end{equation}
    In particular, this term is of order
    
    \begin{equation}\label{estima}
    \sim \frac{12\sqrt{2}}{2^{m}m^{4}}
    \end{equation}
    This asymptotic behavior is very different from those found in (\ref{Taylor}). Hovewer, we can estimate and compare respect to the asymptotics terms in (\ref{Asym}). Expression (\ref{Asy}) can be re-written as  
    
    \begin{equation}\label{Non}
    12\frac{(m-1)![2(m-c-3)]!2!(2c)!2!2!}{(2m)!(m-c-3!)c!1!1!0!}.
    \end{equation}
    The first asymptotic term of this expression, for {\it finite} $c$, is
    
    \begin{equation}
    \frac{3}{2^{1+2c}}\times\frac{(2c)!}{c!}\times\frac{1}{m^{4+c}} \underset{c\to m/2}{=} \frac{3}{2^{m+1}}\times\frac{1}{m^{4}}\times\underbrace{\left[\frac{m}{m}\frac{(m-1)}{m}\frac{(m-2)}{m}\cdots\frac{(m-m/2+1)}{m}\right]}_{f(m)}.
    \end{equation}
    By calling $f(m)$ the term within the big square brackets, we have $$\frac{1}{2^{m/2}}<f(m)\ll 1.$$
    
    We see that, for big $m$,
    
    \begin{equation}
    \frac{3}{2}\times\frac{1}{2^{m}m^{4}}\sim \frac{12\sqrt{2}}{2^{m}m^{4}},
    \end{equation} 
    those factors are of the same order. Thereby, $f(m)$ can be seen as an asymptotic deviation of this centered multinomial term, respect to the non-centered terms expressed in (\ref{Non}).
    
    These centered multinomial contributions are well defined, provided that we consider elements represented by partitions of $b\to m$, such that
    
    \begin{equation}\label{Part}
   \frac{b}{2}+\frac{b}{2};\,\,\, (\frac{b}{2}-1)+\frac{b}{2}+1;\,\,\, (\frac{b}{2}-2)+\frac{b}{2}+1+1;\,\,\, (\frac{b}{2}-1)+(\frac{b}{2}-1)+1+1\cdots
    \end{equation}
    
    Partitions like $(\frac{b}{2}-n)+(\frac{b}{2}+n)$ for finite $n$ generate indeterminate coefficients on the asymptotic expansion (since these partitions have identical first term for $n\, \in \mathbb{N}$ ). Therefore, we only consider partitions like (\ref{Part}). In this case, $b$ is considered even. However, for an arbitrary $b$ we consider
    
    \begin{equation}
    b= \left\lfloor\frac{b}{2}\right\rfloor + \left\lceil\frac{b}{2}\right\rceil,
    \end{equation}
    with $\left\lfloor r\right\rfloor$ ($\left\lceil r\right\rceil$) the nearest integer to rational $r$ from below (above).
    
    The 13 (18, if $m$ is odd) contributing terms in (\ref{h}) for the first four asymptotic terms in the binomial-centered case are in $k=m-1$, $k=m-2$ and $k=m-3$. In particular, following the same procedure in (\ref{hm1}) for these terms, if $m$ is odd, the contribution for $h_{m}^{(1)}$ is
    
    \begin{equation}\label{Bin}
\sim \frac{2}{\sqrt{\pi}}m!m^{\frac{1}{2}}\left[\frac{3\sqrt{2}}{m^{2}}+\frac{67}{2\sqrt{2}m^{3}}+\frac{5763}{16\sqrt{2}m^{4}}+\cdots\right]
    \end{equation}
    
    and if $m$ is even,
    
\begin{equation}
\sim \frac{2}{\sqrt{\pi}}m!m^{\frac{1}{2}}\left[\frac{2\sqrt{2}}{m^{2}}+\frac{21}{\sqrt{2}m^{3}}+\frac{2005}{8\sqrt{2}m^{4}}+\cdots\right]
\end{equation}

The trinomial-centered contribution can be calculated from the partition $$
b= \left\lceil\frac{b}{3}\right\rceil +\left\lfloor\frac{b}{3}\right\rfloor + \left\lceil\frac{b}{3}\right\rceil
$$ for $b\equiv 2\,\,\,\, (\mathrm{mod} \,\,\,3) $, from the partition $$b = \left\lfloor\frac{b}{3}\right\rfloor + \left\lceil\frac{b}{3}\right\rceil+ \left\lfloor\frac{b}{3}\right\rfloor$$ for $b\equiv 1\,\,\,\, (\mathrm{mod} \,\,\,3) $ and, from the partition $$b=\frac{b}{3}+\frac{b}{3}+\frac{b}{3}$$ for  $b\equiv 0\,\,\,\, (\mathrm{mod} \,\,\,3) $, and so on.

It is especially remarkable that these two contributions are negligible respect to expression (\ref{hm1}). Other centered multinomial expansions can be defined from (\ref{Asym}). (In principle, one for each possible multinomial centered multinomial term.) In the same way, they do not seem to give a significant contribution to (\ref{hm1}). Also, unlike the cases in which $k\ll m$, and the multinomial not-centered expansion cases in $k\lesssim m$, the expansions in this $n$-multinomial centered regime depend on $m$ $(\mathrm{mod}\,\,\, n)$. The excelent numerical matching of (\ref{hm1}) with the exact values of $h_{m}^{(1)}$ for large $m$ suggests that, in expression (\ref{h}), it is only necessary to take the limit $k \ll m$ and the multinomial non-centered cases in $k \lesssim m$ to obtain the conventional asymptotic expansion. However, an analysis of this multinomial centered regime and of the family of different asymptotic expansions associated with this regime can be interesting from a mathematical perspective.

\subsection{$N=2$ case}

Recurrence (\ref{Simple}) for the case in which $N=2$ is

\begin{equation}\label{Rec2}
\mathfrak{N}_{m}^{(2)}=2\sum_{j=0}^{m}\binom{m}{j}\mathcal{N}_{c\,j}^{(1)}\mathfrak{N}_{m-j}^{(1)}+ 2\sum_{j=1}^{m}\binom{m}{j}\mathcal{N}_{c\,j}^{(2)}\mathfrak{D}_{m-j}.
\end{equation}

The exact solution of this recurrence for arbitrary $m$ is

\begin{align}\label{N2}
\mathcal{N}_{c\,m}^{(2)}=&\sum_{n=1}^{m}\mathcal{C}_{n}^{m}\left(\frac{\mathfrak{N}_{n}^{(2)}}{2}-\mathfrak{N}_{n}^{(1)}\right) \nonumber \\ &\,\,\, +\sum_{n=1}^{m}\left[2(m-n)-1\right]\mathcal{C}_{n}^{m}\left(\mathfrak{N}_{n}^{(1)}-\mathfrak{D}_{n}\right).
\end{align}
The proof of (\ref{N2}) is obtained by induction. (See Appendix \ref{A1}.) The asymptotic expansion is obtained in a way similar to the case in which $N=1$. Using

\begin{equation}
\frac{\mathfrak{N}_{k}^{(2)}}{2}-\mathfrak{N}_{k}^{(1)}=\left(k+\frac{1}{2}\right)\left(\mathfrak{N}_{k}^{(1)}-\mathfrak{D}_{k}\right),
\end{equation}
we have

\begin{equation}
\mathcal{N}_{c\,m}^{(2)}=\left(\frac{\mathfrak{N}_{m}^{(2)}}{2}-\mathfrak{N}_{m}^{(1)}\right)\frac{2m-1}{2m+1}+\sum_{n=1}^{m-1}\mathcal{C}_{n}^{m}\left(\frac{\mathfrak{N}_{n}^{(2)}}{2}-\mathfrak{N}_{n}^{(1)}\right)\frac{4m-2n-1}{2n+1}.
\end{equation}
Using (\ref{h1}), and after some manipulations, we get

\begin{align}\label{lim}
	h_{m}^{(2)}=& \frac{mm!}{2^{m}}(2m-1)\binom{2m}{m}\left[1+\sum_{n=1}^{m-1}\frac{n}{m}\frac{(4m-2n-1)}{(2m-1)}\frac{(2n)!}{(2m)!}\mathcal{C}_{n}^{m}\right] \nonumber \\  =&\frac{mm!}{2^{m}}(2m-1)\binom{2m}{m}\left[1+\sum_{k=1}^{m-1}\frac{(m-k)}{m}\frac{(2(m+k)-1)}{(2m-1)}\frac{(2[m-k])!}{(2m)!}\mathcal{C}_{m-k}^{m}\right], 
\end{align}
which serves to calculate, respectively, the different limits $n\ll m$ and $k\ll m$, for arbitrarily large $m$. For the same analysis used in case $N=1$, and using expression (\ref{CenBin}), we see that the asymptotic expansion until $a=6$ is

\begin{equation}\label{h2}
h_{m}^{(2)}\sim \frac{1}{\sqrt{\pi}}m!\sqrt{m}(2m-1)2^{m}\left[1-\frac{5}{8m}-\frac{215}{128m^{2}}-\frac{4255}{1024m^{3}}-\frac{627749}{32768m^{4}}-\frac{32650491}{262144m^{5}}-\frac{4251341763}{4194304m^{6}}-\cdots\right].
\end{equation}

Specifically, until $a=6$, for $m=1000$, $m=2000$ and $m=3000$, the aproximation matches with the exact value in the first 17, 19 and 20, digits respectively.

In the same way as in case $N=1$, the multinomial centered dilemma is presented. Hovewer, as in this previous case, it does not seem to have a significant contribution.

The contributing terms in (\ref{lim}) for the first four asymptotic terms in the binomial-centered case are in $k=m-1$, $k=m-2$ and $k=m-3$. In particular, following the same procedure in (\ref{h2}) for these terms, if $m$ is odd, the contribution for $h_{m}^{(2)}$ is

 \begin{equation}
\sim \frac{1}{\sqrt{\pi}}m!m^{\frac{1}{2}}(2m-1)\left[\frac{6\sqrt{2}}{m^{2}}+\frac{32\sqrt{2}}{m^{3}}+\frac{5233}{8\sqrt{2}m^{4}}+\cdots\right]
\end{equation}

and if $m$ is even,

\begin{equation}
\sim \frac{1}{\sqrt{\pi}}m!m^{\frac{1}{2}}(2m-1)\left[\frac{4\sqrt{2}}{m^{2}}+\frac{20\sqrt{2}}{m^{3}}+\frac{1815}{4\sqrt{2}m^{4}}+\cdots\right]
\end{equation}

We tried to find an explicit solution for the recurrence case $N=3$, in terms of the symbols $\mathcal{C}_{n}^{m}$. Although it is possible to find a solution for the first orders, these solutions do not seem to have a simple form that can be generalized for an arbitrary order of $m$, as in cases $N=1$ and $N=2$. (See expressions (\ref{firstCountFormula}) and (\ref{N2}).) The same problem happens for $N>3$, which makes it difficult to obtain a generalized solution of recurrence (\ref{Simple}) for arbitrary $N$. The apparent reason for this is that, for $N\ge 3$, when inserting a tentative solution for $\mathcal{N}_{c\, n}^{(3)}$ (written in terms of the symbols $\mathcal{C}_{r}^{n}$, for $r\le n<m$) in the recurrence (\ref{Simple}), for the case $N=3$, we cannot get a rearrangement of the terms in such way as obtain $\mathcal{N}_{c\, m}^{(3)}$ in terms of $\mathcal{C}_{r}^{m}$. The situation is worse for $N>3$.  This means that our method is not easily extensible for $N\ge 3$.

\section{Discussion and perspectives} \label{D}

In this work, by using simple combinatorial arguments, we have proved a general recurrence formula for the number of different $m$-order Wick contractions that generates connected Feynman diagrams with an arbitrary number of external legs for the fermionic non-relativistic interacing gas. In this case, the recurrence determines the different number of connected Feynman diagrams. The recurrence is easy to process computationally, and it is possible to find exact numerical solutions for a large number of cases in only a few minutes. Other possibilities (self energies, 1pI, skeletons diagrams, etc.,) were not considered in our study, they would certainly need another  approach . However, it would be of great interest to extend our methods for those different cases.

An exact solution is obtained for the cases in which $N=1$ and $N=2$, enabling computation of many terms in the asymptotic expansion for the number of Feynman diagrams in large orders, in these specific cases.

Within this zero-dimensional approach \cite{Cvitanovic} \cite{Borinsky}, the asymptotic analysis is reasonably well understood. The contribution of our work is to provide an equivalent formulation, related to the field-operator approach, showing that this simpler machinery may also contribute with interesting insight. Our analysis shows that, apparently cumbersome expressions as (\ref{Combin2}) can contain relevant combinatorial information and, in particular, a relation with numerical partitions and compositions. Also, our asymptotic analysis for cases $N=1$ and $N=2$ comes from (\ref{Combin2}), in what we call the non-centered multinomial limit. Multinomial centered terms are other types of contributions which seem to be negligible respect to the non-centered contribution. Hovewer, we can define an expansion for every possible multinomial centered contribution. This defines a family of asymptotic expansions related to our problem. The non-centered limit studied here is of great interest as a new asymptotic method that enables the derivation of the same asymptotic expansion calculated by other methods\cite{Cvitanovic}\cite{Borinsky}\cite{Argyres} up to the desired precision order. We hope that our work brings some new features and perspectives in the realm of zero-dimension QFT realm. 

It is as well important to notice that, due to the bijection between Feynman diagrams and the $N$-rooted maps\cite{Gopala2}, our enumerative study is also valid for the $N$-rooted maps. In particular, it would be interesting to study whether our work bears some relation to the generalized catalan numbers used in Ref.\cite{Gopala} in the context of the Eynard-Orantin topological recursion, which is another method for enumeration of Feynman diagrams.   

\section*{ACKNOWLEDGMENTS}

The authors thank the Brazilian agency CNPq (Conselho Nacional de Desenvolvimento Cient\'ifico e Tecnol\'ogico) for partial financial support.   

\bibliography{RefsArticle}

\appendix

\section{Calculus of the asymptotics until $a=6$}\label{A2}

In this appendix, we write explicitly the terms of (\ref{h}) and (\ref{lim}) that contribute to the asymptotic expansion until $a=6$, for the case in which $N=1$.

In the limit where $n \lesssim m$ (or $k \ll m$), only six terms contribute on the left-hand side of (\ref{q}). They are

\begin{align}
-&\frac{2(m-1)}{2m(2m-1)}
-\frac{4(m-2)}{2m(2m-1)(2m-3)}
-\frac{20(m-3)}{2m(2m-1)(2m-3)(2m-5)}-\frac{148(m-4)}{2m(2m-1)(2m-3)(2m-5)(2m-7)} \nonumber \\
&-\frac{1412(m-5)}{2m(2m-1)(2m-3)(2m-5)(2m-7)(2m-9)}- \frac{16324(m-6)}{2m(2m-1)(2m-3)(2m-5)(2m-7)(2m-9)(2m-11)}. \nonumber
\end{align}

 In the limit where $k \lesssim m$, the cases in which $k=m-1,m-2,\cdots,m-5$ are the only ones that present a contribution. For $k=m-1$, all the contributions are given in (\ref{Expan}). Similar expressions are found for the other cases. Summing them up, the total contribution  in this regime is
 
  \begin{align}
  	-&\frac{2}{2m(2m-1)}
  	-\frac{8}{2m(2m-1)(2m-3)}
  	-\frac{60}{2m(2m-1)(2m-3)(2m-5)}-\frac{592}{2m(2m-1)(2m-3)(2m-5)(2m-7)} \nonumber \\
  	&-\frac{7060}{2m(2m-1)(2m-3)(2m-5)(2m-7)(2m-9)}. \nonumber
  \end{align}

For the binomial centered contribution expansion in $N=1$, the cases in which $k=m-1,m-2,m-3$ present contribution in the first three asymptotics terms of (\ref{Bin}). In particular, when $m$ is odd, by adding all the contributions we have

\begin{align}
&\frac{192\left[(m-3)!\right]^{2}(m-1)!}{\left[\left(\frac{m-3}{2}\right)!\right]^{2}(2m)!}+\frac{384(m-5)!\left[(m-1)!\right]^{2}}{\left(\frac{m-5}{2}\right)!\left(\frac{m-1}{2}\right)!(2m)!}+\frac{24(m-3)!\left[(m-1)!\right]^{2}}{\left(\frac{m-3}{2}\right)!\left(\frac{m-1}{2}\right)!(2m)!}+\frac{2\left[(m-3)!\right]^{3}}{\left[\left(\frac{m-1}{2}\right)!\right]^{2}(2m)!}\nonumber\\ &+\frac{384(m-7)!(m-1)!(m+1)!}{\left(\frac{m-7}{2}\right)!\left(\frac{m+1}{2}\right)!(2m)!}+\frac{24(m-7)!(m-1)!(m+1)!}{\left(\frac{m-7}{2}\right)!\left(\frac{m+1}{2}\right)!(2m)!}+\frac{4(m-3)!(m-1)!(m+1)!}{\left(\frac{m-3}{2}\right)!\left(\frac{m+1}{2}\right)!(2m)!}
\end{align} 
and, for $m$ even we have
\begin{align}
\frac{4(m-2)!m!(m-1)!}{\left(\frac{m-2}{2}\right)!\left(\frac{m}{2}\right)!(2m)!}+\frac{12\left[(m-2)!\right]^{2}(m-1)!}{\left[\left(\frac{m-2}{2}\right)!\right]^{2}(2m)!}+\frac{24(m-4)!m!(m-1)!}{\left(\frac{m-4}{2}\right)!\left(\frac{m}{2}\right)!(2m)!}&+\frac{384(m-6)!m!(m-1)!}{\left(\frac{m-6}{2}\right)!\left(\frac{m}{2}\right)!(2m)!}\nonumber \\ &+\frac{384(m-4)!(m-2)!(m-1)!}{\left(\frac{m-4}{2}\right)!\left(\frac{m-2}{2}\right)!(2m)!}
\end{align}	 

\subsection{Case $N=2$}

For the case in which $N=2$, we almost have the same contribution terms. The square bracket terms of (\ref{lim}) and  (\ref{h}) only differ by the multiplicative factor $$\frac{2(m+k)-1}{2m-1}$$ and contribute to exactly the same terms with the new corresponding multiplicative factor. We should also point out that the factor $2/\sqrt{\pi}$ must be replaced by $(2m-1)/\sqrt{\pi}$.

\section{Proof of the recurrence solution in case $N=2$}\label{A1}

For $m=1,2,$ and $3,$ it is easy to see that (\ref{N2}) is satisfied. Suppose that (\ref{N2}) is valid for $k\leq m$. From (\ref{Rec2}), we have that, for $m+1$,

\begin{equation}
\mathcal{N}_{c\,m+1}^{(2)}=\left(\frac{\mathfrak{N}_{m+1}^{(2)}}{2}-\mathfrak{N}_{m+1}^{(1)}\right)-\sum_{j=1}^{m}\binom{m+1}{j}\mathfrak{N}_{m+1-j}^{(1)}\mathcal{N}_{c\,j}^{(1)}-\sum_{j=1}^{m}\binom{m+1}{j}\mathfrak{D}_{m+1-j}\mathcal{N}_{c\,j}^{(2)}-\mathcal{N}_{c\,m+1}^{(1)}.
\end{equation} 

In the third sum, $1\leq j \leq m$. Therefore, we can insert (\ref{N2}) in $\mathcal{N}_{c\,j}^{(2)}$, obtaining the next two terms:

\begin{align}\label{form}
\mathcal{N}_{c\,m+1}^{(2)}=&\left[\left(\frac{\mathfrak{N}_{m+1}^{(2)}}{2}-\mathfrak{N}_{m+1}^{(1)}\right)-\sum_{j=1}^{m}\sum_{n=1}^{j}\binom{m+1}{m+1-j}\mathfrak{D}_{m+1-j}^{(1)}\mathcal{C}_{n}^{j}\left(\frac{\mathfrak{N}_{n}^{(2)}}{2}-\mathfrak{N}_{n}^{(1)}\right)\right]\nonumber\\ &+\left[-\sum_{j=1}^{m}\sum_{n=1}^{j}\left[2(j-n)-1\right]\binom{m+1}{m+1-j}\mathfrak{D}_{m+1-j}\mathcal{C}_{n}^{j}\left(\mathfrak{N}_{n}^{(1)}-\mathfrak{D}_{n}\right)\right. \nonumber \\ &\,\,\,\,\,\,\,\,\,\,\left.-\sum_{j=1}^{m}\sum_{n=1}^{j}\left[2(m+1-j)+1\right]\binom{m+1}{m+1-j}\mathfrak{D}_{m+1-j}\mathcal{C}_{n}^{j}\left(\mathfrak{N}_{n}^{(1)}-\mathfrak{D}_{n}\right)-\mathcal{N}_{c\,m+1}^{(1)}\right].
\end{align}

The procedure used in Ref.\cite{Castro1} to prove case $N=1$ (see formulas (22), (23) and (24) in this reference) can be used for the first term of (\ref{form}), which is identical to

\begin{equation}
\sum_{n=1}^{m+1}\mathcal{C}_{n}^{m+1}\left(\frac{\mathfrak{N}_{n}^{(2)}}{2}-\mathfrak{N}_{n}^{(1)}\right).
\end{equation}

Let us now focus on the second term of (\ref{form}). A careful appreciation lets us find that the first double sum can be rewritten as

\begin{equation}
-\sum_{k=1}^{m}\left[\mathfrak{N}_{k}^{(1)}-\mathfrak{D}_{k}\right]\sum_{n=k}^{m}\left[2(n-k)-1\right]\binom{m+1}{n}\mathfrak{D}_{m+1-n}\mathcal{C}_{k}^{n}.
\end{equation}

On the other hand, the second double sum can be rewritten as

\begin{equation}
-\sum_{k=1}^{m}\left[\mathfrak{N}_{k}^{(1)}-\mathfrak{D}_{k}\right]\sum_{n=k}^{m}\left[2(m+1-n)+1\right]\binom{m+1}{n}\mathfrak{D}_{m+1-n}\mathcal{C}_{k}^{n}.
\end{equation}
By Adding the two previous equations, we obtain

\begin{equation}
-\sum_{k=1}^{m}\left[2(m+1-k)\right]\left[\mathfrak{N}_{k}^{(1)}-\mathfrak{D}_{k}\right]\sum_{n=k}^{m}\binom{m+1}{m+1-n}\mathfrak{D}_{m+1-n}\mathcal{C}_{k}^{n}.
\end{equation}
Note that the dependence of $n$ on factor $2(m+1-k)$ disappeared and the same procedure used in Ref.\cite{Castro1} is valid. (Formulas (23) and (24) of this reference.) Therefore, the previous expression is
 
\begin{equation}
\sum_{k=1}^{m}\left[2(m+1-k)\right]\mathcal{C}_{k}^{m+1}\left(\mathfrak{N}_{k}^{(1)}-\mathfrak{D}_{k}\right)
\end{equation}
and, by using (\ref{firstCountFormula}) for $\mathcal{N}_{c\,m+1}^{(1)}$ in (\ref{form}), we have

\begin{align}
	\mathcal{N}_{c\,m+1}^{(2)}=&\sum_{n=1}^{m+1}\mathcal{C}_{n}^{m+1}\left(\frac{\mathfrak{N}_{n}^{(2)}}{2}-\mathfrak{N}_{n}^{(1)}\right) \nonumber \\ &\,\,\, +\sum_{n=1}^{m+1}\left[2(m+1-n)-1\right]\mathcal{C}_{n}^{m+1}\left(\mathfrak{N}_{n}^{(1)}-\mathfrak{D}_{n}\right),
\end{align}
which proves (\ref{N2}).

\section{Generating functions and recurrences}\label{A3}

An alternative way to derive relations (\ref{Total}) and (\ref{Simple}) is using generating functions. Let 

\begin{equation}
F(x,y)=\sum_{N=0}^{\infty}\sum_{m=0}^{\infty}\frac{\mathfrak{N}_{m}^{(N)}}{\left(N!\right)^{2}m!}x^{N}y^{m}
\end{equation}
and 

\begin{equation}
G(x,y)=Log\left(\sum_{m=0}^{\infty}\frac{\mathfrak{D}_{m}}{m!}y^{m}\right)+\sum_{n=1}^{\infty}\sum_{m=n-1}^{\infty}\frac{\mathcal{N}_{c\,m}^{(n)}}{n!m!}x^{n}y^{m}
\end{equation}
be the generating functions of $\mathfrak{N}_{m}^{(N)}$ and $\mathcal{N}_{c\,m}^{(n)}$, respectively. Connected and disconnected generating functions of Feynman diagrams are easily related. This relation is maintained in zero-diemnsional field theory. In particular,

\begin{equation}\label{F}
F(x,y)=\exp\left(G(x,y)\right)
\end{equation}

\begin{equation}
\sum_{N=0}^{\infty}\sum_{m=0}^{\infty}\frac{\mathfrak{N}_{m}^{(N)}}{\left(N!\right)^{2}m!}x^{N}y^{m}=\sum_{m=0}^{\infty}\frac{\mathfrak{D}_{m}}{m!}y^{m}\left( 1+\sum_{n=1}^{\infty}\sum_{m=n-1}^{\infty}\frac{\mathcal{N}_{c\,m}^{(n)}}{n!m!}x^{n}y^{m}+\cdots+\frac{1}{l!}\left[\sum_{n=1}^{\infty}\sum_{m=n-1}^{\infty}\frac{\mathcal{N}_{c\,m}^{(n)}}{n!m!}x^{n}y^{m}\right]^{l}+\cdots\right).
\end{equation}

By redefining the sum indices on the right side, using the multinomial theorem, and carefully comparing term by term, we obtain expression (\ref{Total}).

Formula (\ref{Simple}) is obtained by differentiating (\ref{F}) consecutively and evaluating in $x=0$, defining

\begin{equation}
\frac{dF}{dx}(x,y)=F^{\prime}(x,y)
\end{equation}

From expression (\ref{F}), we get all the special cases of expression (\ref{Simple}). In particular, we obtain case $N=1$ from

\begin{equation}
F^{\prime}(0,y)=F(0,y)G^{\prime}(0,y),
\end{equation}

$N=2$ from 

\begin{equation}
F^{\prime\prime}(0,y)=F(0,y)G^{\prime\prime}(0,y)+F^{\prime}(0,y)G^{\prime}(0,y),
\end{equation}

$N=3$ from

\begin{equation}
F^{\prime\prime\prime}(0,y)=F(0,y)G^{\prime\prime\prime}(0,y)+2F^{\prime}(0,y)G^{\prime\prime}(0,y)+F^{\prime\prime}(0,y)G^{\prime}(0,y),
\end{equation}

and so on.

\end{document}